\documentstyle[times,emulateapj]{article}

\lefthead{Rector, Stocke \& Perlman}
\righthead{A Search for Low-Luminosity BL Lacs}

\def\vvmax{$\langle V/V_{max} \rangle$}

\begin{document}

\title{A Search for Low-Luminosity BL Lacertae Objects}
\author{Travis A. Rector\altaffilmark{1,2,3,4} and John T.
Stocke\altaffilmark{1,2,3}}
\affil{Center for Astrophysics and Space Astronomy, University of Colorado, Boulder, Colorado 80309-0389}
\authoremail{rector@casa.Colorado.EDU}

\and

\author{Eric S. Perlman}
\affil{Space Telescope Science Institute, Baltimore, MD 21218}

\altaffiltext{1}{Visiting Astronomer, Kitt Peak National Observatory, National
Optical Astronomy Observatories, which is operated by the Association of
Universities for Research in Astronomy, Inc. (AURA) under cooperative agreement
with the National Science Foundation.}
\altaffiltext{2}{Visiting Astronomer, Multiple Mirror Telescope Observatory.  MMTO
is owned and operated by the Smithsonian Astrophysical Observatory and the
University of Arizona.}
\altaffiltext{3}{Visiting Astronomer, National Radio Astronomy Observatory.  NRAO
is a facility of the National Science Foundation operated under cooperative
agreement by Associated Universities, Inc.}
\altaffiltext{4}{Current address:  National Optical Astronomy Observatories, 950 N. Cherry Ave., Tucson, AZ
85719}





\begin{abstract}
Many properties of BL Lacs have become explicable in terms of the ``relativistic
beaming'' hypothesis whereby BL Lacs are FR-1 radio galaxies viewed nearly along
the jet axis. However, a possible problem with this model is that a transition
population between beamed BL Lacs and  unbeamed FR-1s has not been detected.  A
transition population of ``low-luminosity BL Lacs'' was predicted to exist in
abundance in X-ray-selected samples such as the {\it Einstein} Extended Medium
Sensitivity Survey (EMSS) by Browne \& March\~a.  However, these BL Lacs may
have been misidentified as clusters of galaxies.  We have conducted a search for
such objects in the EMSS with the ROSAT HRI; and here we present ROSAT HRI
images, optical spectra and VLA radio maps for a small number of BL Lacs which
were previously misidentified in the EMSS catalog as clusters of galaxies. 
While these objects are slightly lower in luminosity than other EMSS BL Lacs,
their properties are too similar to the other BL Lacs in the EMSS sample to
``bridge the gap'' between BL Lacs and FR-1 radio galaxies.  Also, the number of
new BL Lacs found are too few to alter significantly the X-ray luminosity
function or \vvmax\ value for the X-ray-selected EMSS BL Lac sample. Thus, these
observations do not explain fully the \vvmax\ discrepancy between the X-ray and
radio-selected BL Lac samples.
\end{abstract}


\keywords{BL Lacertae Objects --- AGN --- Unification Models}


%

\section{Introduction}

BL Lacertae Objects are a very enigmatic type of active galactic nucleus (AGN).
They are highly luminous (e.g., log $L_x =$ 44-46 ergs s$^{-1}$; Morris et al.
1991, hereafter M91), dominated by a non-thermal, nearly-featureless continuum
(which often makes redshift determinations difficult) and exhibit several
characteristics indicative of relativistic outflows nearly along the line of sight 
(e.g., Doppler-boosted luminosity: Padovani \& Urry 1990; apparent superluminal
motion of VLBI radio components: Zensus 1989; and variability on timescales as
short as minutes to hours: Heidt \& Wagner 1996).  Indeed, BL Lacs display some of
the most extreme behavior of any class of AGN. Given their large X-ray luminosities
BL Lacs are found in modestly large numbers in X-ray surveys (e.g., 43 in the {\it
Einstein} Extended Medium Sensitivity Survey (EMSS): Stocke et al.  1991,
Rector et al. 1999; 66 in the {\it Einstein} ``Slew Survey": Perlman et al. 1996b;
and 120 in the ROSAT-Green Bank (RGB) Catalog: Laurent-Muehleisen et al. 1998). 
There may well be of order 1000 BL Lacs hidden among all the sources in the ROSAT
All-Sky Survey (RASS; e.g., Nass et al. 1996; Bade et al. 1998).

Many observations (e.g., extended radio luminosity and morphology: Antonucci \&
Ulvestad 1985; Perlman \& Stocke 1993; host galaxy luminosity and morphology:
Abraham et al. 1991; Wurtz, Stocke \& Yee 1996; comparative space densities and
luminosity functions: Padovani and Urry 1990, M91)  suggest that BL Lac objects are
relativistically-beamed ``Fanaroff \& Riley class 1'' (FR-1) radio galaxies, which
are low-luminosity, edge-darkened sources often found in rich and poor cluster
environments. BL Lacs discovered by both X-ray-selection (XBLs) and radio-selection
(RBLs) techniques share these characteristics, strongly suggesting that XBLs and
RBLs are from the same parent population. And although XBLs and RBLs have somewhat
different observed properties (e.g., XBLs have larger starlight and smaller
polarized flux fractions optically: M91; Jannuzi et al. 1993, 1994 and lower radio core
dominance values: Perlman \& Stocke 1993, Laurent-Muehleisen et al. 1993), most of 
these differences can be
accomodated either by invoking different mean viewing angles, wherein XBLs are
viewed further from their beaming axis (e.g., Stocke et al. 1989) or by invoking
different electron energy distributions (Giommi \& Padovani 1994). But many RBLs
have extended radio and optical emission-line luminosities inconsistent with being
FR-1s (Rector \& Stocke 1999) and, most suprising, Stocke \& Rector (1997) have 
found a large excess of
MgII absorption systems in the 1Jy RBL sample, suggestive of gravitational
lensing.  Thus, while XBLs seem mostly consistent with being beamed FR-1s, RBLs
seem to be a mixture of beamed FR-1s and FR-2s as well as a few gravitationally-lensed
quasars.

Recent work on the clustering environments of BL Lacs by Wurtz et al. (1997)
finds that BL Lacs avoid rich clusters and the most luminous galaxies (i.e., the
brightest cluster galaxies) at low redshifts, suggesting that there may be a
problem with this simple picture.  Also, Owen, Ledlow \& Keel (1996) failed to
find substantial numbers of BL Lac candidates among their large ($\sim 200$)
sample of rich cluster radio galaxies.  Either there exists some physical reason
why BL Lacs avoid rich clusters at low $z$ or current surveys systematically miss
BL Lacs in rich clusters (see Wurtz et al. 1997 for suggested reasons).  Another
observation that does not fit the simple beamed FR-1 picture is the disparity in
\vvmax\ values for RBLs and XBLs ($0.61 \pm 0.05$ for RBLs in the 1Jy survey;
Rector \& Stocke 1999; $0.33 \pm0.06$ for XBLs in the EMSS; M91, Perlman et al.
1996b), which suggests that they may be from different populations and/or that one
or both samples are incomplete.  Indeed, Della Ceca (1993) first pointed out that
raising the X-ray flux limit of the EMSS sample to $f_x \geq 10^{-12}$ ergs s$^{-1}$
cm$^{-2}$ raises \vvmax\ to $0.48 \pm 0.06$, which is consistent with no evolution. 
Thus incompleteness of the EMSS sample at faint fluxes is a real concern.

Browne \& March\~a (1993; hereafter BM93) and March\~a \& Browne (1995; hereafter
MB95) describe an effect which can make radio- and X-ray-selected BL Lacs difficult
to recognize optically, thus introducing the possibility that such samples are
incomplete:  A BL Lac object at low redshift with a flux density near the survey
limit (i.e., a ``low-luminosity BL Lac") would be difficult to identify optically
because the weak nonthermal continuum from the AGN could be fainter than the
starlight from the luminous host elliptical galaxy, thus causing these X-ray or
radio sources to be misidentified as either ``normal'' elliptical galaxies or as
clusters of galaxies rather than as BL Lacs.  This effect (hereafter called the
``B-M effect") will result in a paucity of BL Lacs at low $z$, thus flattening the
observed X-ray luminosity function (XLF) at the low-luminosity end (MB95).  The
\vvmax\ for the sample will also be artificially decreased, as this effect causes
the flux limit of the sample to be underestimated.  BM93 suggested that the EMSS BL
Lac sample may be incomplete on the basis of the sample flux limit and the XLF.  
However, in a more detailed analysis MB95 find that, while there is some evidence
for ``hidden'' BL Lacs, the XLF and \vvmax\ statistics of the EMSS sample cannot be
altered significantly by this selection effect; and so it probably cannot explain
the \vvmax\ discrepancy between the EMSS and 1Jy samples (MB95).

In this paper we present evidence for a small population of low-luminosity BL Lac objects 
that were originally misidentified as clusters of galaxies in the EMSS (\S
2). These objects were identified using ROSAT HRI observations which reveal the
presence of point-like X-ray emission at the location of a low-power radio galaxy,
thus revealing the presence of a weak AGN (\S 3).  Subsequent optical spectroscopy
and VLA imaging confirm that these sources are BL Lacs (\S 4).  Individual sources
are discussed in \S 5.  Since the EMSS is the only statistically-complete
X-ray-selected sample of BL Lacs for which the XLF and evolution have been
calculated, these results affect the XLF and evolution determinations as well as
physical models of BL Lacs (e.g., Ghisellini \& Maraschi 1989).  However, based upon
the small number of new EMSS BL Lacs discovered, the modifications to earlier XLF
evolution and \vvmax\ determinations are slight (\S 6.) 

\section{The Sample Selected for ROSAT HRI Observation}

The {\it Einstein} Medium Sensitivity Survey (EMSS; Gioia et al. 1990; Stocke et al.
1991; Maccacaro et al. 1994) contains 835 faint ( $> 7$ x $10^{-14}$ ergs cm$^{-2}$
s$^{-1}$ in the $0.3-3.5$ keV soft X-ray band) X-ray sources discovered
serendipitously with {\it Einstein} Imaging Proportional Counter (IPC) images
obtained of various targets at high Galactic latitude ($b > 20$\arcdeg).  To be
included in the EMSS a source detection limit of 4$\sigma$ above the X-ray plus detector 
backgrounds is imposed. 

The full EMSS sample of BL Lac objects currently consists of 43 objects with
X-ray fluxes from $10^{-11}$ to $10^{-13}$ ergs s$^{-1}$ cm$^{-2}$ in the 0.3-3.5
keV band (Rector et al. 1999).  The identification of these objects followed both a
spectroscopic and a photometric set of criteria as outlined in Stocke et al. (1991),
as  did the identification of the other classes of X-ray sources identified in the
EMSS (e.g., QSOs and Seyferts, clusters of galaxies, normal galaxies and Galactic
stars).

Based upon the nearly complete identifications of sources in the EMSS survey, we
assembled a list of the best candidates of potentially misidentified clusters of
galaxies. In particular, we concentrated our efforts on the EMSS subsample from
which the ``complete'' BL Lac sample of M91 was extracted.  This sample of BL Lacs
was selected by the following criteria:  An X-ray flux of $f_x \geq 5.0$ x
$10^{-13}$ ergs s$^{-1}$ cm$^{-2}$ and a declination limit of $\delta \geq
-20\arcdeg$.  The X-ray-selected objects drawn from the EMSS using these constraints
are fully identified; thus incompleteness of the M91 BL Lac sample can only occur from
the misidentification of sources. The sample selected for ROSAT HRI observations includes
all suspect sources within the bounds of the M91 subsample plus the best cases for
misidentified BL Lacs in the rest of the EMSS. Our criteria for what constitutes a suspect
source are detailed below.

\subsection{Candidates from the M91 Subsample}

In compiling the suspect source list we eliminated very well-resolved X-ray sources whose
extended flux so dominates over any possibly embedded point source that it would
be below the flux limit of the EMSS.  Many of the EMSS clusters have been studied with
ROSAT, confirming their extended nature. So, using either ROSAT PSPC or HRI or the
original {\it Einstein} IPC images, the maximum point source contribution to the IPC detect
cell counts is measured by fitting a point source at the location of the the detected 
cluster radio galaxy or galaxies. Since most EMSS sources were detected quite close to the
signal-to-noise limit (4$\sigma$), it was unlikely that any point source contribution
to extended cluster emission would be detected on its own. But,
it is important to note that, while these radio galaxies might be excluded from being EMSS BL Lacs,
they still could be low-luminosity BL Lacs; i.e., the cluster is the correct X-ray
identification, but a low-luminosity BL Lac is nevertheless present within the cluster. One 
such source (MS 0011.7+0837) was discovered in the course of this study.

X-ray clusters lacking radio
sources were also eliminated from the suspect list, 
as radio-quiet BL Lacs are not believed to exist (Stocke et al.
1990). VLA surveys of EMSS clusters (Stocke et al. 1991, Stocke et al. 1999) have been conducted, which
are deep enough to
detect most FR-1 radio galaxies in a cluster.  The VLA snapshot survey of all EMSS sources reported in
Stocke et al. (1991) has a 5$\sigma$ detection limit of $\sim$ 1 mJy at 5 GHz, which should be sufficient to
detect radio galaxies in nearby clusters (a detection limit of log $L_r \geq 23$ W Hz$^{-1}$ for $z \leq
0.2$). A deeper VLA survey has recently been completed at 1.4 GHz for all EMSS clusters with z $\geq$ 0.3
(Stocke et al. 1999). These new observations were timed to achieve a point-source radio power limit of log
$L_r \geq$ 23.5 W Hz$^{-1}$. While this limit is not quite deep enough to detect all FR-1s,
BL Lacs have Doppler-boosted cores which add significantly to the total radio power level
of an FR-1. Thus, the combination of these two sets of radio observations should be adequate to
detect all potential BL Lacs in EMSS clusters.
Using the observed optical luminosities of BL Lac host galaxies from Wurtz, Stocke
\& Yee (1996), the power limits of the above mentioned VLA surveys require that any non-detection
falls at $\alpha_{ro} \leq$ 0.2, well below the level of BL Lacs (see Stocke et al. 1991 and Section 
6 herein). Therefore, we have eliminated clusters from the suspect list if no radio sources 
were detected by these surveys.

Clusters with radio galaxies whose optical spectra are
dominated by bright emission lines (i.e., ``cooling flow'' clusters) were originally eliminated
from the suspect list due to the usual BL Lac classification criterion which requires
``featureless'' optical spectra (see below).  However, the luminous, low-ionization  optical emission 
lines found in the spectra of the brightest cluster galaxies (BCGs) in cooling flow clusters are very
extended and
not directly related to either the broad- or narrow-line regions of an AGN.
Moreover, these clusters also contain central
``cusps'' in their X-ray surface brightness distributions, which could hide the presence of a
low-luminosity point source associated with the BCG and these BCGs are
often weak radio emitters. Thus, these BCGs are potential low-luminosity BL Lac objects if we ignore the
optical line emission. So, although the
presence of an X-ray point source is not the usual interpretation for the X-ray ``cusps'' 
in cooling flow clusters (e.g., Fabian 1994), 
we have scrutinized all the cooling flow clusters whose BCGs are radio galaxies
to ensure that the excess central surface brightness
could not hide a point source so luminous that a potential EMSS BL Lac could be present (see above).
Eleven such clusters were scrutinized using archival ROSAT and/or Einstein data. All but four
of these clusters are easily too extended to harbor a point source luminous enough
to be a member of the EMSS sample. Two of these four (MS 0102.3+3255 \& MS 1244.2+7114) are 
marginally resolved by current
observations so that a BL Lac could be present. Another cooling flow cluster (MS
1455.0+2232) is very well-resolved by a pointed PSPC observation. But, because this source was
detected at a very high significance level in the EMSS, an embedded point source of only $\sim$ 20\%
of the IPC detect cell flux would still be an EMSS source. Since the cooling flow ``cusp''
in the PSPC image of this source could hide a point source which is $\sim$ 15\% of the detect
cell flux, we must consider the BCG in MS 1455.0+2232 a possible low-luminosity
BL Lac.  The final source (MS 2348.0+2913)
appears unresolved both in the ROSAT all-sky survey and in an off-axis PSPC pointing and so
must be considered a likely BL Lac object. However, since we had not considered the
possibility that ``cooling flow clusters'' could contain BL Lacs until this point was raised by
the referee, we were unable to obtain ROSAT HRI images of these interesting sources.
 
No EMSS sources identified as ``normal''
galaxies meet the above criteria to be suspect; most normal galaxies have quite extended X-ray
emission and only a few are radio emitters.  We note that deep ROSAT HRI
observations of normal galaxies from the EMSS catalog by ourselves and others
(Dahlem \& Stuhrmann 1998; Stocke, Rector \& Griffiths 1999) do not find galaxies
with strong, nuclear point sources.  Any point sources present do not dominate the total
X-ray flux; thus these objects were not detected as BL Lacs on the basis of their
X-ray emission and so would not be included in the EMSS BL Lac sample.  

A total of
seven candidates for misidentification emerged from this selection process, all of
which have been observed  by the ROSAT HRI (excepting the four cooling flow clusters identified
as suspect only after the HRI observations were obtained). These are all of the sources in the
EMSS subsample of M91 that meet the above criteria.


\subsection{Other Low-Luminosity BL Lac Candidates \& Classification Criteria}

Because it is possible that low-luminosity BL Lacs may be present in the EMSS but
not in the bright M91 complete subsample, we have also included five additional
EMSS sources which meet the selection criteria to be suspect identifications but are below the flux or
declination limits of the M91 sample. These five may not be all of the suspect sources
in the EMSS sample which are outside the M91 subsample, but are just the best cases we have
found to this date.  All of these objects have been observed
by the ROSAT HRI as well.


Besides these new HRI observations, the classification criteria previously used to
classify an X-ray source as a BL Lac require both optical spectroscopy and radio
imaging. Therefore, we obtained new optical and radio data to determine whether
these new candidates meet these BL Lac criteria.
But, because these new sources may stretch the boundaries of the BL Lac class, 
the previously used criteria may need modification. These criteria are 
(Stocke et al. 1991):   
(1) unresolved at X-ray
wavelengths; (2) comparable to or more luminous than a typical FR-1 radio  galaxy at
radio wavelengths (e.g., log $L_r \geq 23$ W Hz$^{-1}$). (3) ``featureless'' optical
spectra in terms of emission lines (i.e., any emission line must be quite weak; i.e.,
$W_{\lambda} \leq 5$\AA); and (4) ``featureless'' optical spectra in terms of
absorption lines (i.e., the presence of a nonthermal component to the optical
continuum, as indicated by a CaII break ``contrast'' of $< 25\%$; which corresponds
to $D(4000) \leq 1.33$ as defined in Dressler \& Shectman 1987).  The last two
criteria are somewhat arbitrary as originally described by Stocke et al. (1991). 
Specifically, criterion \#4 is violated by three objects herein which are otherwise BL
Lac-like, as BM93 predicted.  Further, March\~a et al. (1996) proposed a further relaxation
of the criteria, which is discussed in \S 6.  Also because some low-luminosity BL Lacs may be
embedded in extended cluster X-ray emission, criterion \#1 may need reevaluation as well. 
For sources observed with the VLA, an
edge-darkened FR-1 extended radio luminosity and morphology surrounding a bright
point source is expected but is not required for classifying the source as a BL Lac, since a few
BL Lacs have FR-2 morphologies. Also implicit in the above criteria is that the X-ray 
and radio core positions must coincide.  We
note that optical polarization, often a criterion used for classification of
BL Lacs, is not considered here primarily because polarimetry has not yet been
obtained for these objects.  However, polarimetry blueward of the CaII break would
provide important confirmation that the present candidates are BL Lac objects.

A secondary set of photometric criteria were also used in Stocke et al. (1991) to
cross-check potential identifications.  Specifically for BL Lac objects, a
broad-band color criterion is made possible by their unique spectral
energy distributions.  That is, XBLs are found to occupy a nearly unique area in
the ($\alpha_{ro},\alpha_{ox}$) plot (see \S 4), where these spectral
indices refer to radio-to-optical and optical-to-X-ray respectively as defined in
Stocke et al. (1991).  This criterion was rigorously tested by using the {\it
Einstein} Slew Survey data to predict BL Lac object identifications (Perlman et al.
1996a) and was found to be $>90$\% successful in predicting such identifications
prior to optical spectroscopy.  As with previously identified EMSS BL Lacs, we will
use this photometric criterion to check low-luminosity BL Lac candidates.

\section{X-Ray Observations}

Twelve BL Lac candidates were observed with the ROSAT HRI during AO1, AO5 and AO7. 
The log of ROSAT observations for our sample are listed in Table~\ref{tbl-1}.  Col.
[1] is the EMSS object name.  Cols. [2]-[4] list the ROSAT observation request
(ROR) number, the date of the pointed observation, and the effective exposure
time; Col [5] indicates whether or not the object meets the EMSS complete subsample
criterion of M91 described above; and Cols [6] and [7] list the HRI position of the
source; for extended sources the position marks the peak flux.

\begin{figure}
\vspace{6.0in}
\plotfiddle{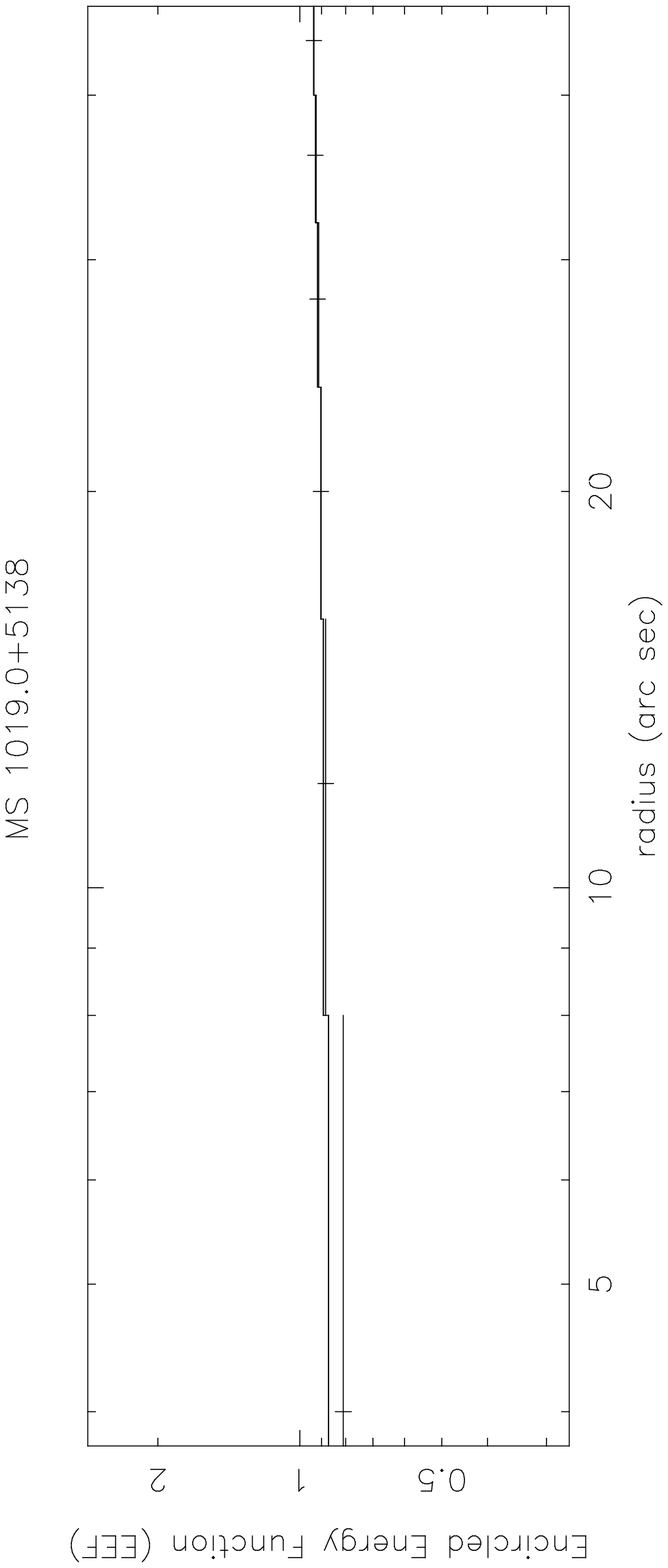}{0in}{-90.}{32.}{32.}{-125}{420}
\plotfiddle{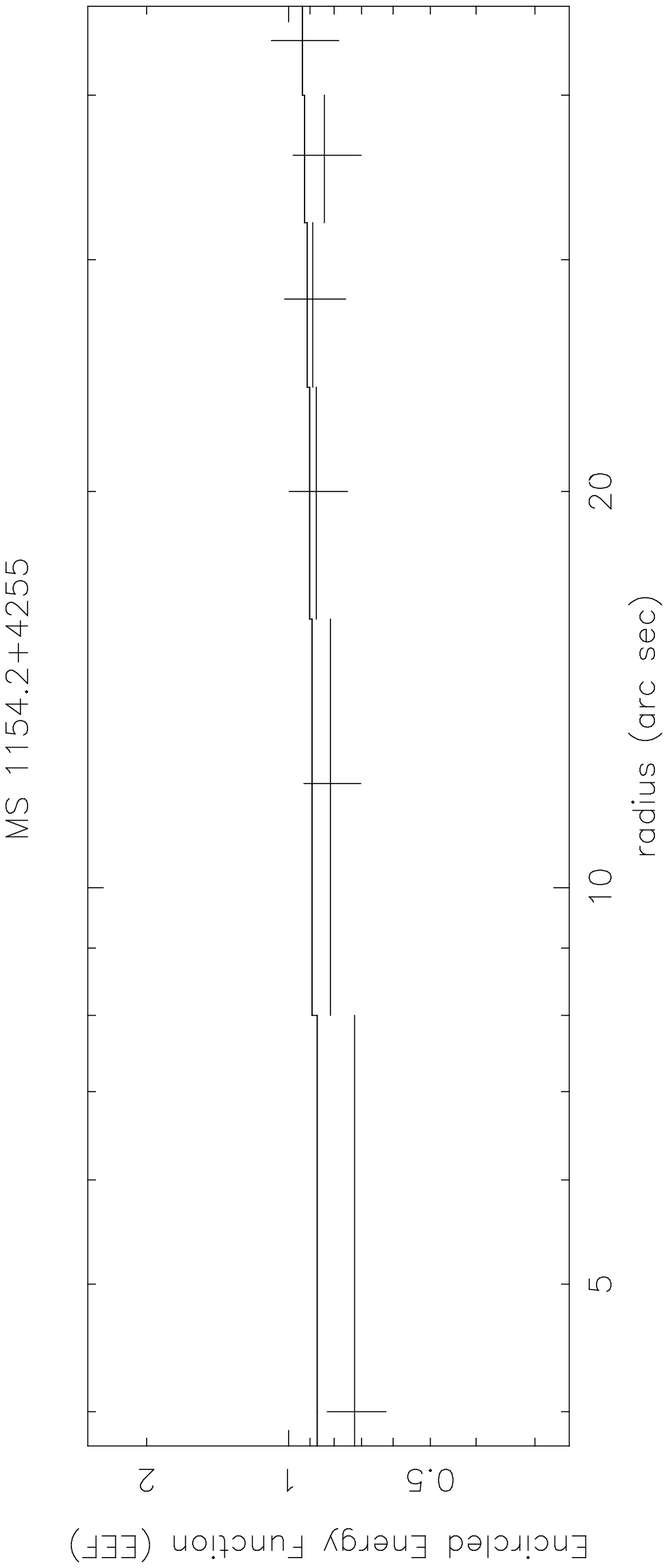}{0in}{-90.}{32.}{32.}{-125}{340}
\plotfiddle{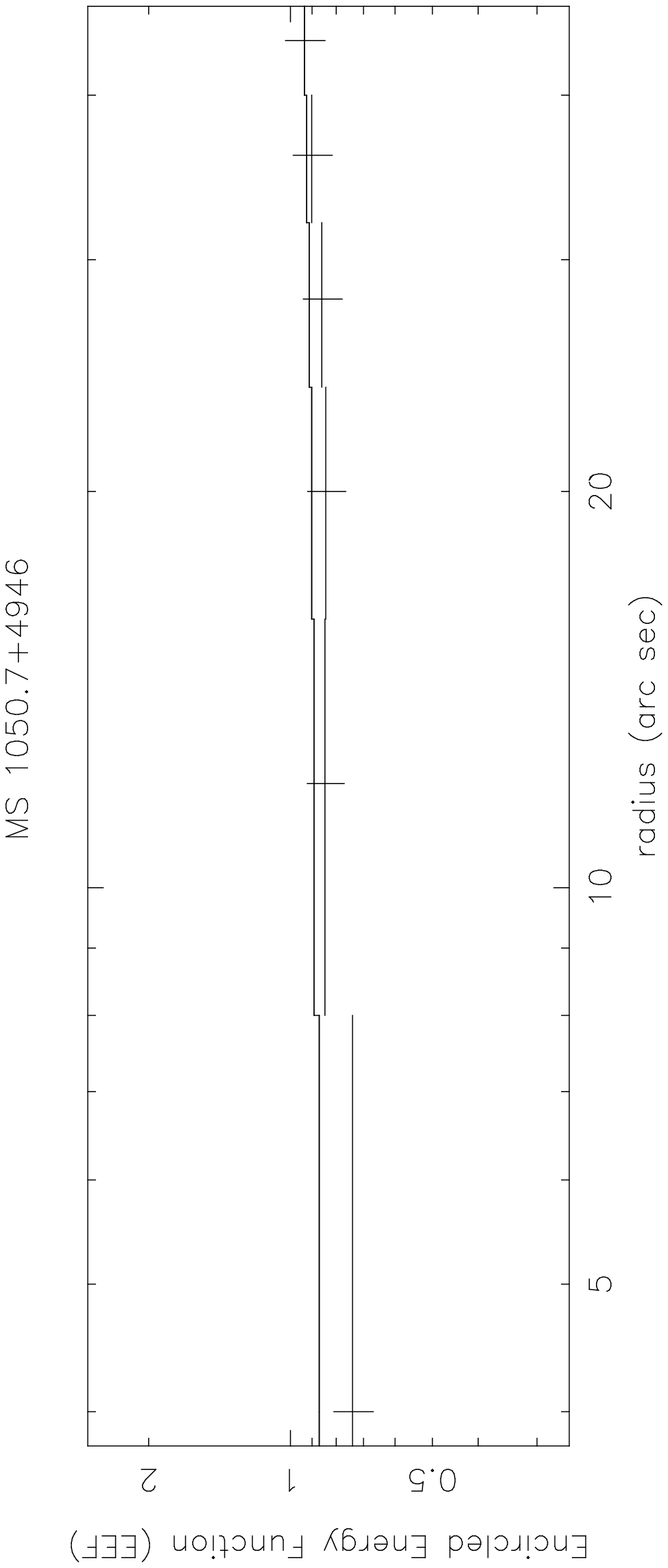}{0in}{-90.}{32.}{32.}{-125}{260}
\plotfiddle{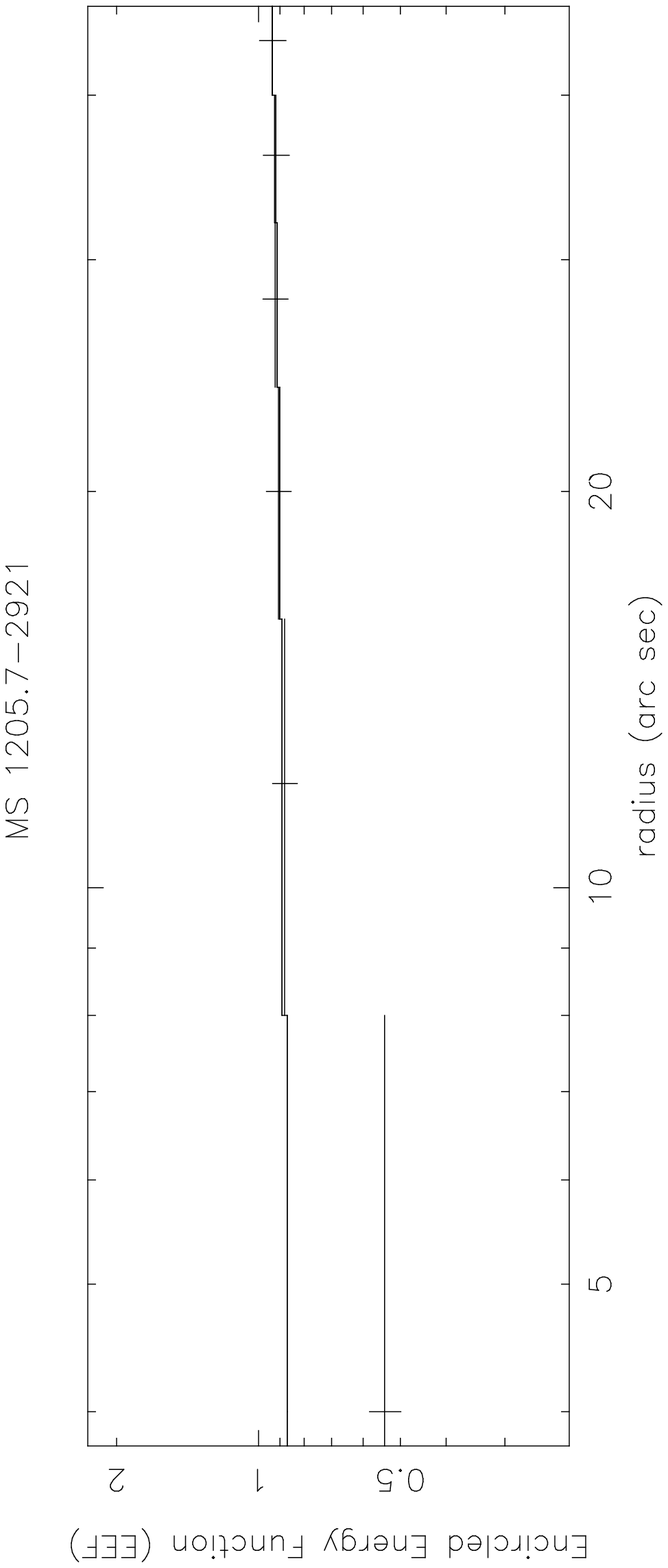}{0in}{-90.}{32.}{32.}{-125}{180}
\plotfiddle{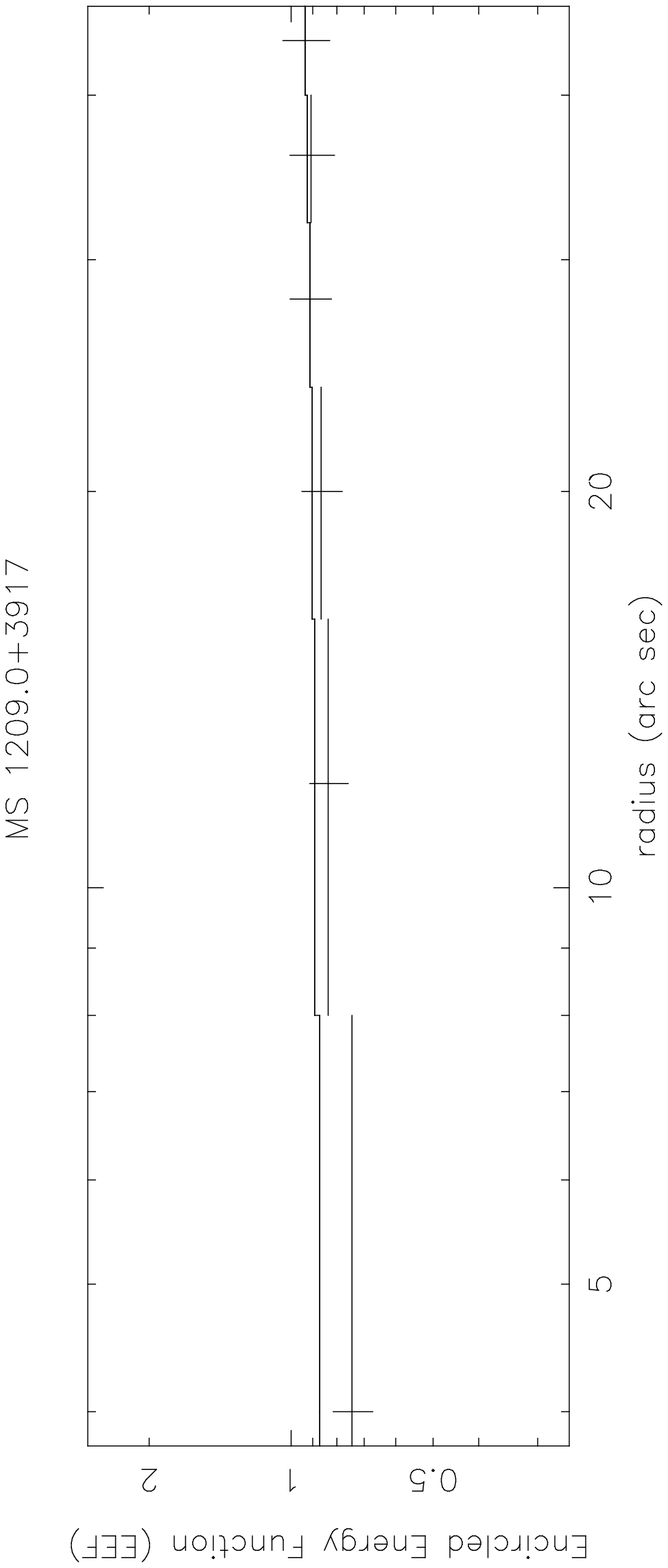}{0in}{-90.}{32.}{32.}{-125}{100}

\caption{The radial energy profile (data points with 1$\sigma$ error bars) for the
unresolved sources plotted against the Encircled Energy Funtion (EEF; solid lines)
for the ROSAT HRI. The object's EEF is normalized to the HRI point-source EEF at a
radius of $50\arcsec$. Note that each source does not match exactly the EEF at a
radius of $\leq 8\arcsec$.  This is due to an ``excess halo'' at $\sim 10\arcsec$,
which is also present for stars within the field; so that these objects are
completely consistent with being unresolved sources. 
\label{fig1}}
\end{figure}
\begin{figure}
\vspace{5.0in}
\plotfiddle{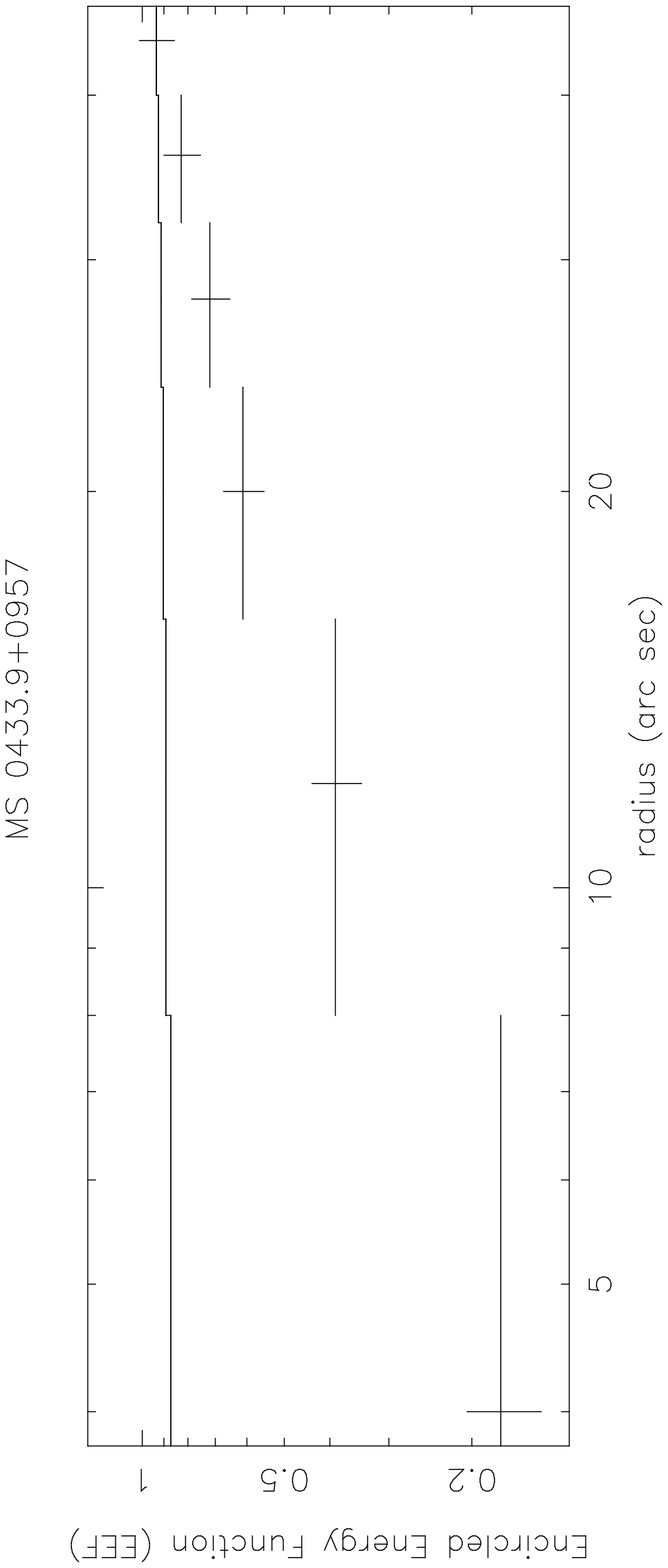}{0in}{-90.}{32.}{32.}{-125}{300}
\plotfiddle{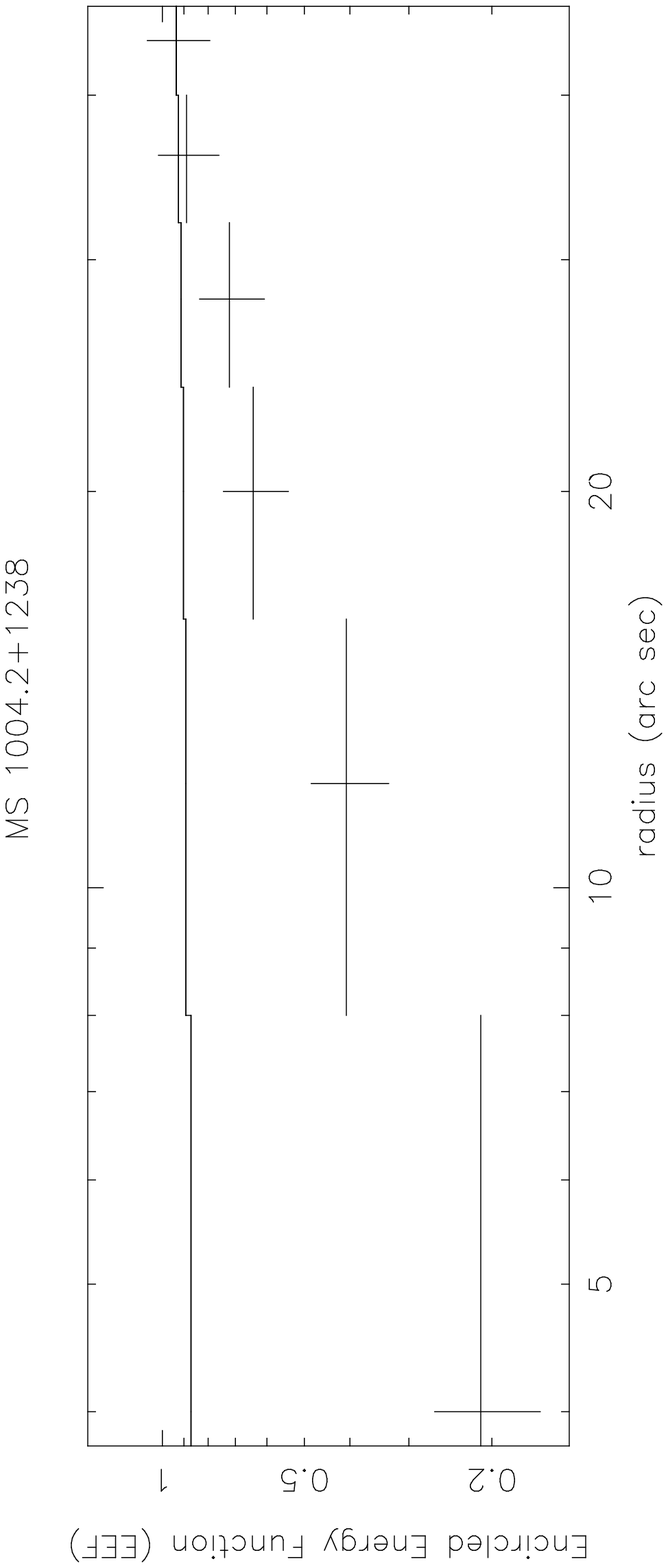}{0in}{-90.}{32.}{32.}{-125}{200}
\plotfiddle{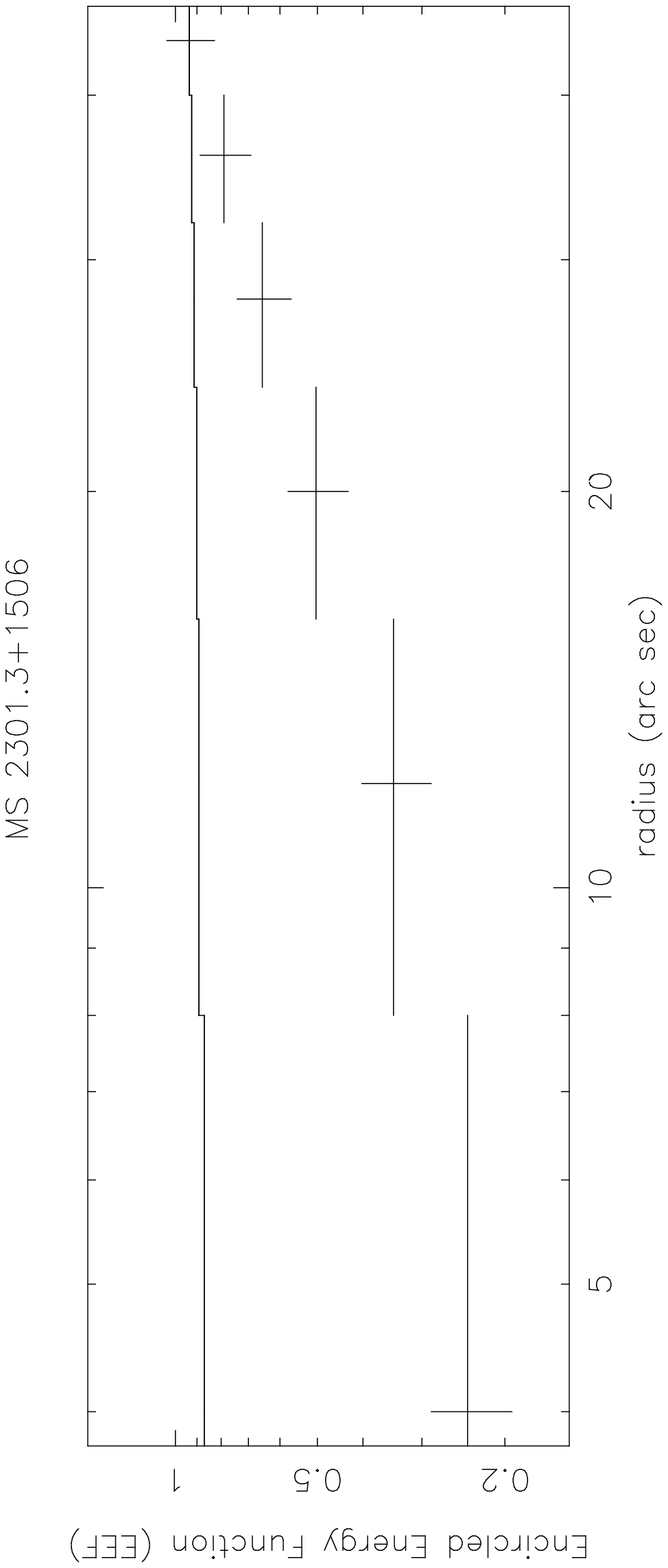}{0in}{-90.}{32.}{32.}{-125}{100}

\caption{The radial energy profile (data points with 1$\sigma$ error bars) for the
resolved sources plotted against the Encircled Energy Funtion (EEF; solid lines)
for the ROSAT HRI (see Figure~\ref{fig1} for a full description of the EEF).  Each
object's total flux is dominated by the extended flux, thus causing the energy
profile to lie well below the HRI EEF at small and large radii, not just at $\leq
8\arcsec$ where the objects in Figure~\ref{fig1} exhibit the ``excess halo." 
\label{fig2}}
\end{figure}

XIMAGE (v2.5; part of the XANADU\footnote{XANADU is provided by LHEA/NASA.}
package) and FTOOLS\footnote{FTOOLS is provided by LHEA/NASA.} (v.4.0) were used to
determine whether or not a source was resolved by matching the nominal on-axis
Encircled Energy Function (EEF) of the HRI to the radial profile of each source. 
The EEF should be viewed as the HRI response function to a point source; however,
other effects (e.g., pointing ``jitter" and aspect solution errors; Morse 1994) may
alter the actual point source response from the EEF.  The radial profiles for the
unresolved and resolved sources are shown in Figures~\ref{fig1} and~\ref{fig2}
respectively.  Radial profiles for MS 1317.0-2111 and MS 1520.1+3002 are not shown
because no X-ray source was detected by the HRI at the position of the radio galaxy
for either source.  Six sources are unambiguously unresolved, one (MS 0011.7+0837)
is marginally resolved and the remaining are well-resolved by these observations. 
Figure~\ref{fig1} shows that even the unresolved sources do not exactly match the
the EEF at a radius of $\leq 8\arcsec$, whereas 87\% of the point-source flux
should be within this radius.  This is due to a halo in excess of the nominal EEF
at a radius of $\sim 10\arcsec$.  This ``excess halo'' has been noted in other HRI
observational programs, including the calibration star HZ 43, but an explanation
has not yet been found (D.E. Harris 1997, private communication).  
Since the ``excess halo'' is also
consistently present in the point spread function of stars within the fields of our
observations, we are confident that these sources are not resolved despite this
$\sim 1\sigma$ excess over the nominal EEF.  Contoured, grey-scale HRI maps of the
resolved sources are shown in Figure~\ref{fig4}.

Two sources (MS 0011.7+0837 and MS 1826.5+7256) did not exactly match the HRI EEF
and it was questionable as to whether or not they were resolved.  After further
analysis of MS 0011.7+0837 we are confident that it is spatially resolved. To
confirm that its extended flux was not due to misalignment of the individual
observation intervals (``OBIs'') or jitter within each OBI, the following test was
done on the nearby EMSS source MS 0011.6+0840, a bright K0 V star fortuitously
within the field of the HRI and sufficiently ``on axis'' to have the same EEF (its
offset from field center is 4.8\arcmin).  The observation was split into the four
different OBIs and the centroid for MS 0011.6+0840 was determined for each OBI. 
The OBIs were then restacked, with each OBI linearly shifted  to align the
centroids.  MS 0011.7+0837 and MS 0011.6+0840 were then compared to the HRI EEF. 
The original stacking of OBIs suffered from minor misalignment, and there is some
evidence for jitter within two of the four OBIs.  However, the energy profile for
MS 0011.6+0840 extracted from the restacked image matches the nominal HRI EEF quite
well while the energy profile for MS 0011.7+0837 does not (Figure~\ref{fig3}); the
probability that the energy profile of MS 0011.7+0837 is the same as MS 0011.6+0840
is $\sim1$\% (the variance summed over all the annuli is $\sigma = 3.77 \pm 0.27$). 
Further, MS 0011.7+0837 is significantly elongated along the E-W axis (see
Figure~\ref{fig4}).  However, because MS 0011.7+0837 contains either a radio galaxy
or low-luminosity BL Lac embedded in extended cluster emission, we have included it
in the other tests of the unresolved sources for thoroughness (see below).

The HRI field for MS 1826.5+7256 also contains an unidentified star at $\alpha =
18^h27^m42.8^s, \delta = +72\arcdeg54\arcmin53.0\arcsec$ (B1950), which allowed us
to perform the same analysis as above.  The radial energy profiles for neither MS
1826.5+7256 nor the in-field star match the HRI EEF.   However, the radial energy
profile for MS 1826.5+7256 is completely consistent with that of the star; the
probability that they are the same is $> 95$\% ($\sigma = 0.01 \pm 0.34$).  The
cause of the poor fit of MS 1826.5+7256 and the star to the nominal HRI EEF is
unknown, but the integration time for this source was significantly longer than for
the other sources, so it may have been caused by excessive jitter (MS 1826.5+7256
was observed in AO1).

The HRI position of MS 1826.5+7256 coincides with the optical position of a faint
M5 V star (V $\approx 17$) at $\alpha = 18^h26^m25.8^s, \delta =
+72\arcdeg56\arcmin06.5\arcsec$ (B1950).  So MS 1826.5+7256 is unresolved, but it is
not a BL Lac.  Note that the optical position of MS 1826.5+7256 given in Maccacaro
et al. (1994) is in error; the optical position given is that of a K star
$19\arcsec$ to the NE of the true optical identification at the position quoted
above.


The fluxes of the unresolved sources were measured by using an aperture of
75\arcsec\ in radius to determine the source count rate.  As can be seen in 
Figure~\ref{fig1} this radius is more than sufficient to enclose all the flux
from the source.  The background count rate
was determined by using a 100\arcsec\ to 200\arcsec\ annulus around the source; the
background was then subtracted from the source count rate.  The HRI has essentially
no spectral resolution so the counts are modeled using a power-law spectrum of the
form $F_{\nu} \propto \nu^{-\alpha}$.  The X-ray spectral index $\alpha$ was
estimated using the luminosity vs. X-ray spectral index correlation from Padovani
\& Giommi (1996) for all but MS 1019.0+5139 and MS 1209.0+3917.  The spectral index
for these two sources was determined by ROSAT PSPC observations (Blair et al.
1997).  The flux was corrected for neutral Hydrogen absorption, assuming the
Morrison \& Maccammon (1983) absorption model and neutral hydrogen column densities
along the line of sight obtained from Gioia et al. (1990).  Table~\ref{tbl-2}
contains the X-ray properties of the unresolved sources (excepting MS 1826.5+7256,
which is discussed above) and MS 0011.7+0837, including by column: [1] EMSS source
name; [2] whether or not it is a member of the complete sample of M91; [3] the
logarithmic density of galactic neutral Hydrogen toward the source (cm$^{-2}$;
Gioia et al. 1990); [4] the soft X-ray power-law energy spectral index; [5] the
ROSAT HRI flux ($0.1-2.4$ keV) in units of $10^{-13}$ erg s$^{-1}$ cm$^{-2}$,
corrected for the $n_H$ in column 2 and assuming the spectral index in column 4;
[6] is the logarithmic HRI luminosity in erg s$^{-1}$ assuming $H_0 = 50$ km
s$^{-1}$ Mpc$^{-1}$, $q_0 = 0$; [7] the observed EMSS IPC flux ($0.3-3.5$ keV) in
units of $10^{-13}$ erg s$^{-1}$ cm$^{-2}$, corrected for $n_H$ absorption and
assuming a spectral index of
$\alpha_x = 1$  (Gioia et al. (1990); [8] is the same as column [7] except assuming
$\alpha_x$ from column [4]; Column [7] is the best estimate of the IPC flux if the
source is dominated by extended cluster emission, column [8] is the best estimate
if the source is dominated by a BL Lac object; and [9] is the fraction of predicted
to observed IPC flux (see description below).

It is possible that faint, very extended structure from an associated cluster may
be present around some of the unresolved sources. This extended emission may have
been detected by the low-resolution {\it Einstein} IPC but was undetected by the
higher-resolution ROSAT HRI due to a lower signal-to-noise ratio per pixel for the
HRI.
To determine whether or not there was any ``missing flux'' in the HRI observations we
converted the HRI flux to an ``expected'' IPC flux by accounting for the different
bands of the ROSAT HRI (0.1 - 2.4 keV) and the {\it Einstein} IPC (0.3 - 3.5 keV).  
We assume the X-ray spectral index $\alpha_x$ for the BL Lac to be valid across the
{\it Einstein} IPC band, which is reasonable because the HRI and IPC bands
significantly overlap.  The HRI flux measurements are comparable (i.e., within 50\%)
to the IPC flux for three of the five BL Lac candidates.  The exceptions, MS
1019.0+5139 and MS 1205.7-2921, are approximately three times brighter than measured
by the IPC.  Allowing for the variability of BL Lac X-ray emission (factors of $\sim
2-3$ are typical; e.g., Sambruna et al. 1995) we conclude that the HRI flux levels
for the five BL Lac candidates are consistent with the {\it Einstein} IPC flux being
entirely from an unresolved BL Lac X-ray emitter and that significant amounts of
extended diffuse flux are not present. On the other hand, the total detected HRI flux for
MS 0011.7+0837 is only 17\% of the IPC flux.  While it may be possible that this source
is a BL Lac that was observed by the IPC in a flaring state, it is much more likely
that the HRI has simply missed faint extended flux associated with a cluster because
it is marginally resolved by the HRI.  The optical and radio data (see below)
also support the conclusion that the X-ray source MS 0011.7+0837 is not simply a BL Lac object but
either a radio galaxy or low-luminosity BL Lac embedded in diffuse cluster
emission.  Even if MS 0011.7+0837 is a combination of cluster and BL Lac X-ray
emission, the analysis shown in Table~\ref{tbl-2} shows that any X-ray point source
present likely contributes only a small fraction to the IPC detection.  

The HRI maps for three sources, MS 0011.7+0837, MS 1004.2+1238 and MS 2301.3+1506,
do not rule out the possibility of a small but signficiant (20-50\%) contribution to
the total X-ray flux by a BL Lac object embedded within the cluster.  We note that
these objects are \textit{not} excluded from consideration as BL Lacs simply because they are
resolved, but because these sources would not meet the EMSS 4$\sigma$ detection
criterion without the extended X-ray flux.  In fact, no clusters in the EMSS are
sufficiently bright such that an embedded BL Lac with $< 20$\% of the total X-ray
flux could be detected without the associated cluster emission (see \S 2.1).

\section{Optical and Radio Observations}

Optical spectra of the radio galaxy in MS 0011.7+0837 and the BL Lac candidates MS
1050.7+4946 and MS 1154.1+4255 were obtained at the KPNO 2.1m telescope.  A
spectrum of the BL Lac candidate MS 1019.0+5139 was obtained at the Multiple
Mirror Telescope Observatory (MMTO).  Low-resolution gratings were used to obtain
one full spectral order (4000\AA\ - 8000\AA) at $\sim 5$\AA\ resolution.  These
spectra were extracted and analyzed in IRAF and flux calibration was performed using
observations of flux standards at similar airmass.  No correction was made for the
loss of light at the slit; thus the derived fluxes are probably uncertain by a
factor of $\sim 2$.  In the 2.1m spectra the flux calibration redward of 6500\AA\
is corrupted due to second-order overlap and should not be trusted.  
Also, the slit was not rotated to the parallactic angle, and so it is possible that
the spectral shape and therefore the measurement of $D(4000)$ could be slightly
affected.
Archival spectra from the original EMSS identification program for the unresolved
X-ray sources MS 1209.0+3917 and MS 1205.7-2921, taken at the Canada-France-Hawaii
Telescope (CFHT) by Isabella Gioia and at the Las Campanas 100" by Simon Morris
respectively, were also analyzed.  The spectra of MS 1205.7-2921 and of MS
1209.0+3917 are not of high quality; higher SNR spectra should be obtained in order
to determine more accurate values for $D(4000)$ and better emission line limits. 
These spectra are shown in Figure~\ref{fig5}.

The spectra were used to determine whether or not the object met the BL Lac
classification criteria (see \S 1) and to determine an accurate redshift.  The
redshift of the host galaxy for each object was determined by cross correlating the
object spectrum with the giant elliptical galaxy NGC 3245 (as described in 
Ellingson \& Yee 1994); the technique has an accuracy of $\pm 0.001$.  We
note that the redshift reported here for MS 1205.7-2921 ($z=0.249$) is different
than previously reported in Maccacaro et al. (1994).  The CaII break $D(4000)$
value was determined by measuring the continuum flux level on either side of the
CaII break (see Dressler \& Shectman 1987 for the complete formalism).  The typical
$D(4000)$ for a normal cluster elliptical or S0 galaxy is $\approx 2.0$.  Weak
emission lines were detected in two objects, MS 1019.0+5139 and MS 1050.7+4946; the
latter object has H$\alpha$ only marginally detected.  For the other objects upper
limits on the presence of emission lines for [O II], H$\beta$, [O III] and
H$\alpha$ (if within the observed band) were calculated.  The poorest limit for
these lines is listed in Table~\ref{tbl-3}.  Interestingly only two other BL
Lacs (MS 1235.4+6315 and MS 2316.3-4222) in the EMSS sample show any emission line
features, and these are both quite weak [O II] emission lines.  In contrast, a large
fraction of radio-selected BL Lacs in the 1Jy sample (Stickel et al. 1991) possess
weak but luminous emission lines in their spectra (Stickel, Fried \& K\"uhr 1993;
Rector \& Stocke 1999).  Figure~\ref{fig7} shows the location of these new BL Lacs
in the diagnostic plot of Stocke et al. (1991).

Deep VLA continuum maps were taken for two sources which meet the selection
criteria of the M91 complete sample (MS 1019.0+5139 and MS 1050.7+4946; see
Figure~\ref{fig6} for a map of the latter) on 27 April 1997.  We chose to observe
with the B-array at 20cm with a 50MHz bandwidth to maximize sensitivity to
extended, steep-spectrum structure while achieving $\sim 4$\arcsec\ resolution,
which is ideally suited for these nearby sources.  Eight 10-minute scans, each
bracketed by a 90-second scan on a primary VLA phase calibrator, yielded $\sim
150,000$ ``visibilities" for each source.  Scans were widely spaced to optimize
$(u,v)$ coverage; achieving a 0.02 mJy beam$^{-1}$ ($1\sigma$ RMS) noise level.
Additionally a 20cm A-array ``snapshot" was taken of MS 0011.7+0837 (also shown in
Figure~\ref{fig6}), which has a 1$\sigma$ noise level of 0.24 mJy beam$^{-1}$.  

The data were reduced in a standard manner with the NRAO AIPS software package. 
The core flux was determined by fitting a point source to the core.  The extended
flux is the total flux, as determined by the region around the source
containing all the extended structure, minus the core flux. The standard K and
bandpass corrections are applied to the luminosity calculations: a power-law
continuum of the form $F_{\nu} \propto \nu^{+\alpha}$ is assumed where $\alpha =
-0.8$ for extended flux and $\alpha = +0.3$ for the core (values typical of BL Lacs;
Perlman \& Stocke 1993; Stickel et al. 1991).  We note that Perlman \& Stocke (1993)
instead use $\alpha = -0.3$ for the core (a value more typical of FR-1s).
However, the effect of this discrepancy is slight
because EMSS BL Lacs are typically at low redshift ($z \approx 0.3$).  Perlman \& Stocke
(1993) describe additional cosmological distance effects which can bias the measured core
and extended fluxes, particularly so in FR-1 sources at high redshift ($z > 0.2$). 
Corrections for these biases are not applied because they are not significant at the
redshifts of MS 0011.7+0837 and MS 1050.7+4946 and are strongly source dependent.

The morphologies of MS 0011.7+0837 and MS 1050.7+4946 (shown in Figure~\ref{fig6})
are typical of low-power FR-1 sources excepting that MS 1050.7+4946 is somewhat more
core dominated than a typical FR-1.  Its core dominance is comparable to other EMSS
BL Lacs; however, based upon the VLA configuration used and its low $z$ it is
possible that not all the extended flux was detected.  Thus the extended flux and
core dominance values reported in Table~\ref{tbl-3} should be considered as upper
limits.  

Given its modest redshift of $z=0.141$ it is unusual that no extended flux
was detected for MS 1019.0+5139.  We quote a very conservative upper limit on the
extended flux in this source, which assumes an intrinsically large source (100 x 30
kpc$^2$) with an extended surface brightness just below the noise level per beam;
however this limit is still very low for an FR-1 and is lower than any other BL Lac
in the EMSS sample (Rector et al. 1999).  
MS 1019.0+5139 was detected by both the Faint Images of the Radio Sky at Twenty
Centimeters (FIRST; B-array at 20cm; White
et al. 1997) and the NRAO VLA Sky Survey
(NVSS; D-array at 20cm; Condon et al. 1998) surveys but was
not resolved in either. 
Interestingly, the flux of this source in the FIRST survey catalog is 3.3 mJy, nearly
identical to the flux measured in our map; however the NVSS flux is significantly
higher ($5.1 \pm 0.5$ mJy), suggesting that extended flux may have been detected on
larger scales; although variability is also a possible explanation.  
A high dynamic-range C-configuration
20cm map should be make of this source to determine whether or not it has significant
extended flux.

The optical and radio properties of the unresolved sources are given in
Table~\ref{tbl-3}.  Columns include: [1] EMSS source name; [2] redshift; [3] the
V-band magnitude (Stocke et al. 1991); [4] the equivalent width of the strongest
emission feature (or the largest $3\sigma$ upper limit if none are present) [5] the
CaII break value $D(4000)$, as defined in Dressler \& Shectman (1987); [6] and [7]
the VLA core and total radio flux at 20cm (mJy), except where noted; [8] the core
dominance, defined as the core to extended flux ratio; and [9] and [10] the VLA
20cm core and extended luminosities in W Hz$^{-1}$.

The optical spectra for the newly discovered EMSS BL Lacs are shown in
Figure~\ref{fig5}.  The spectrum for a typical EMSS BL Lac can be described as that
of a luminous elliptical galaxy with a significant nonthermal component blueward of
the CaII H\&K break (rest $\lambda \approx 4000$\AA).  It should be clear from
Figure~\ref{fig5} that starlight from the host elliptical dominates the nonthermal
continuum from the BL Lac in the V-band ($\sim 5500$\AA).  Since, in these cases
the X-ray and radio flux is dominated by the BL Lac but the optical flux is
dominated by the host galaxy, the ($\alpha_{ox}, \alpha_{ro}$) values do not
accurately reflect the properties of the BL Lacs themselves.  To correct for this
effect we determined the optical flux for each BL Lac by the following method:  The
host elliptical is assumed to have a CaII break strength of 50\% (i.e., $D(4000) =
2.0$).  Flux in excess of this amount blueward of the CaII break is assumed to be
from the BL Lac itself.  This flux is extrapolated to 5500\AA\ by using the
spectral slope measured directly blueward of the break, if possible.  If the
spectrum is insufficient to determine the slope an optical spectral index of
$\alpha_o = 1.0$ is assumed (Stocke et al. 1991).  All of the BL Lacs in the EMSS
sample with a significant elliptical component in their spectrum (i.e., $D(4000) >
1.0$) are corrected in this manner in Figure~\ref{fig8}.  Since this correction
removes optical flux, the ($\alpha_{ox}, \alpha_{ro}$) values for the BL Lacs are
displaced upwards and to the left.
We note that we do not correct the measured equivalent width of emission lines for
the presence of the host galaxy.  As pointed out by March\~a et al. (1996), quite
strong emission lines ($W_{\lambda} \leq 50$\AA\ relative to the AGN alone) could be
present without exceeding the traditional BL Lac criterion ($W_{\lambda} \leq 5$\AA\
relative to the AGN \textit{plus} host galaxy continuum).

We specifically note that MS 1019.0+5139 has a position in Figure~\ref{fig8}
($\alpha_{ox} = 0.72; \alpha_{ro} = 0.41$) fully consistent with a BL Lac object and
not with a radio-quiet QSO or Seyfert galaxy.  We are concerned about the nature of
this particular object because it is the only EMSS BL Lac that possesses weak
optical emission lines, a flat X-ray spectral slope and an unresolved radio
source.  Although Seyferts have much stronger emission lines, the other two
attributes of MS 1019.0+5139 are more characteristic of Seyferts than BL Lacs. 
However the radio core power is much more characteristic of a radio galaxy than of
a Seyfert.  This potential transitional object deserves further observational
attention.

\section{Discussion of Individual Sources}

\begin{figure}
\vspace{7.0in}

\plotfiddle{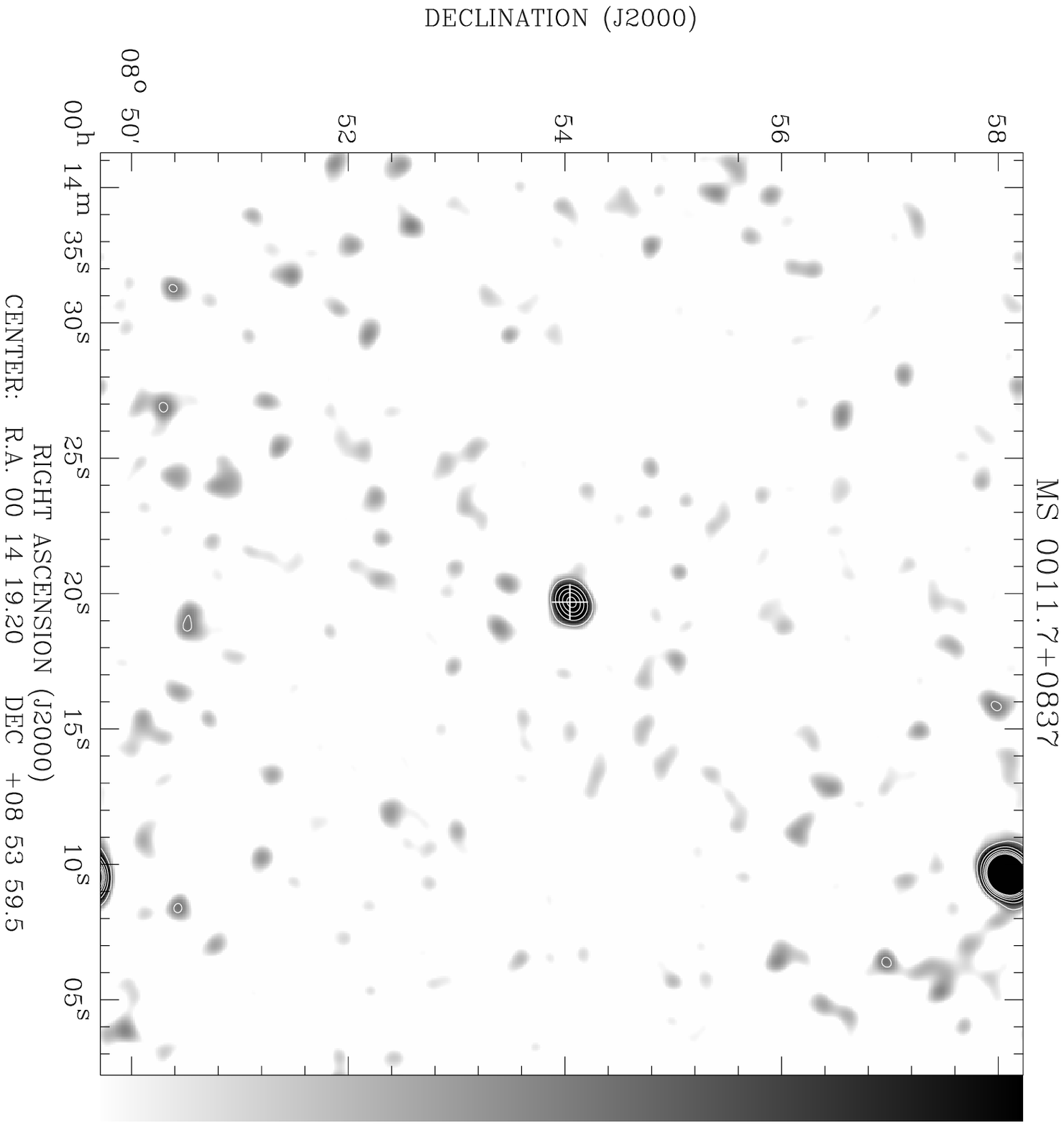}{0in}{90.}{40.}{40.}{80}{300}
\plotfiddle{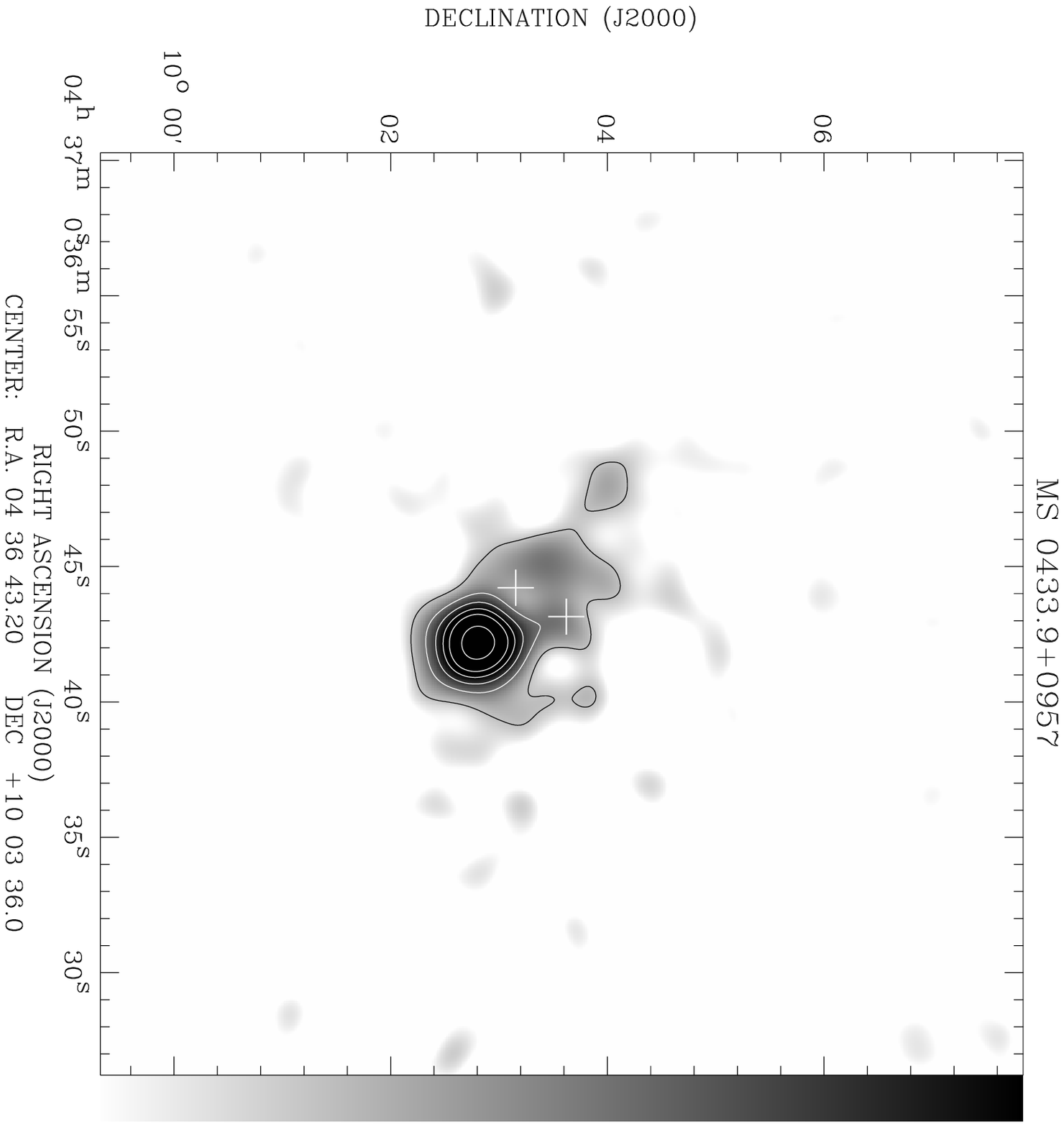}{0in}{90.}{40.}{40.}{300}{321}
\plotfiddle{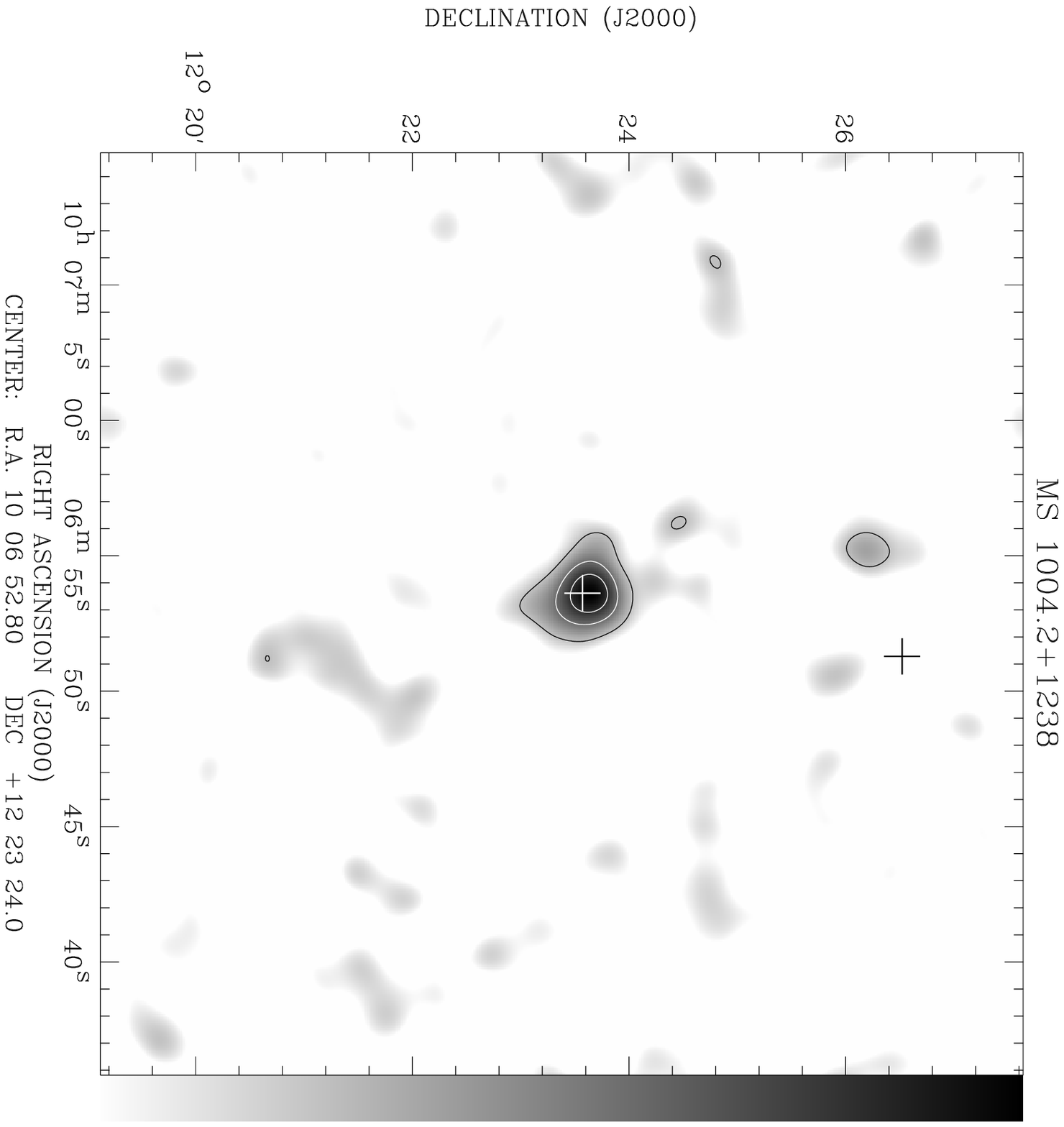}{0in}{90.}{40.}{40.}{80}{140}
\plotfiddle{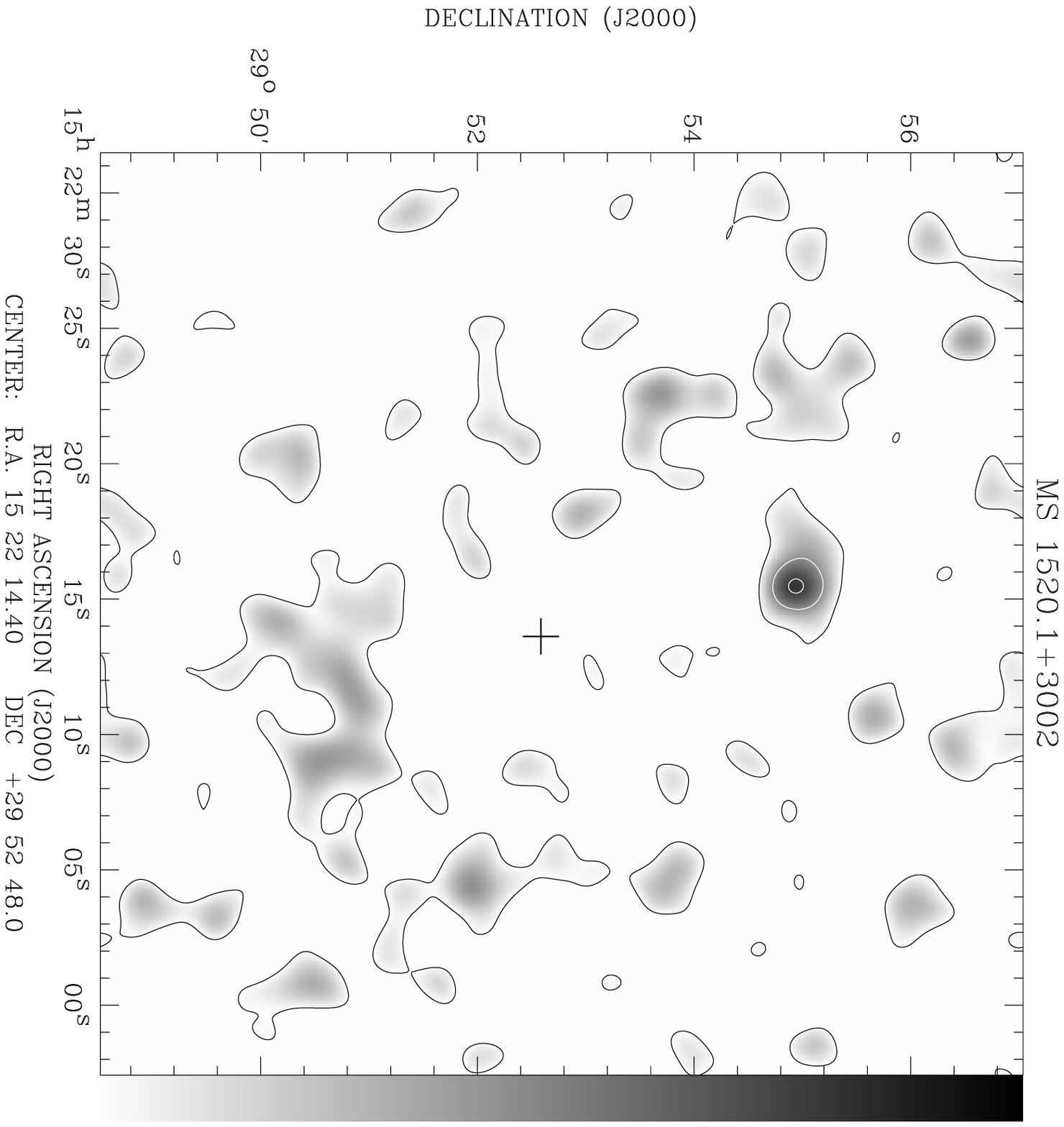}{0in}{90.}{40.}{40.}{300}{161}
\plotfiddle{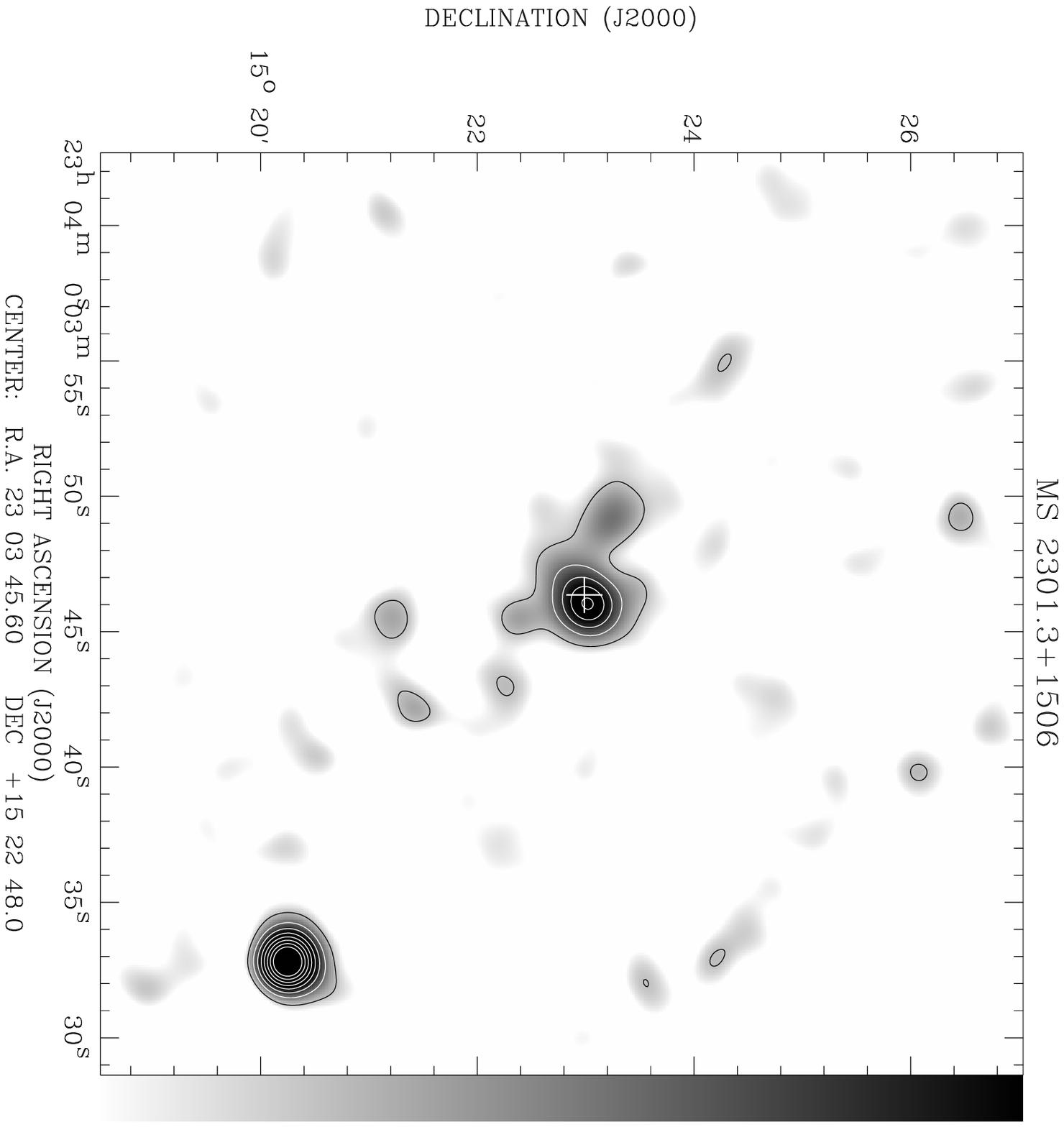}{0in}{90.}{40.}{40.}{190}{-20}

\caption{Contoured, logarithmic grey-scale X-ray maps of the resolved sources.  For
all sources the greyscale is logarithmic from 0.015 to 0.05 counts arcsec$^{-1}$. 
The map for MS 0011.7+0837 is smoothed by a Gaussian kernel equivalent to the on-axis
point-spread funtion (PSF) of the HRI.  The other maps are smoothed by a Gaussian
twice the width of the HRI PSF to enhance the extended flux.  The contour levels
for the map of MS 0011.7+0837 are linearly increasing from 0.03 in 0.03 increments
(in counts arcsec$^{-1}$).  For the other maps the contours linearly increase from
0.02 in 0.01 increments. The positions of radio galaxies within the field are
marked with crosses.  Further discussion of individual maps can be found in \S 5.}

\label{fig4}
\end{figure}

\begin{figure}
\vspace{5.0in}
\plotfiddle{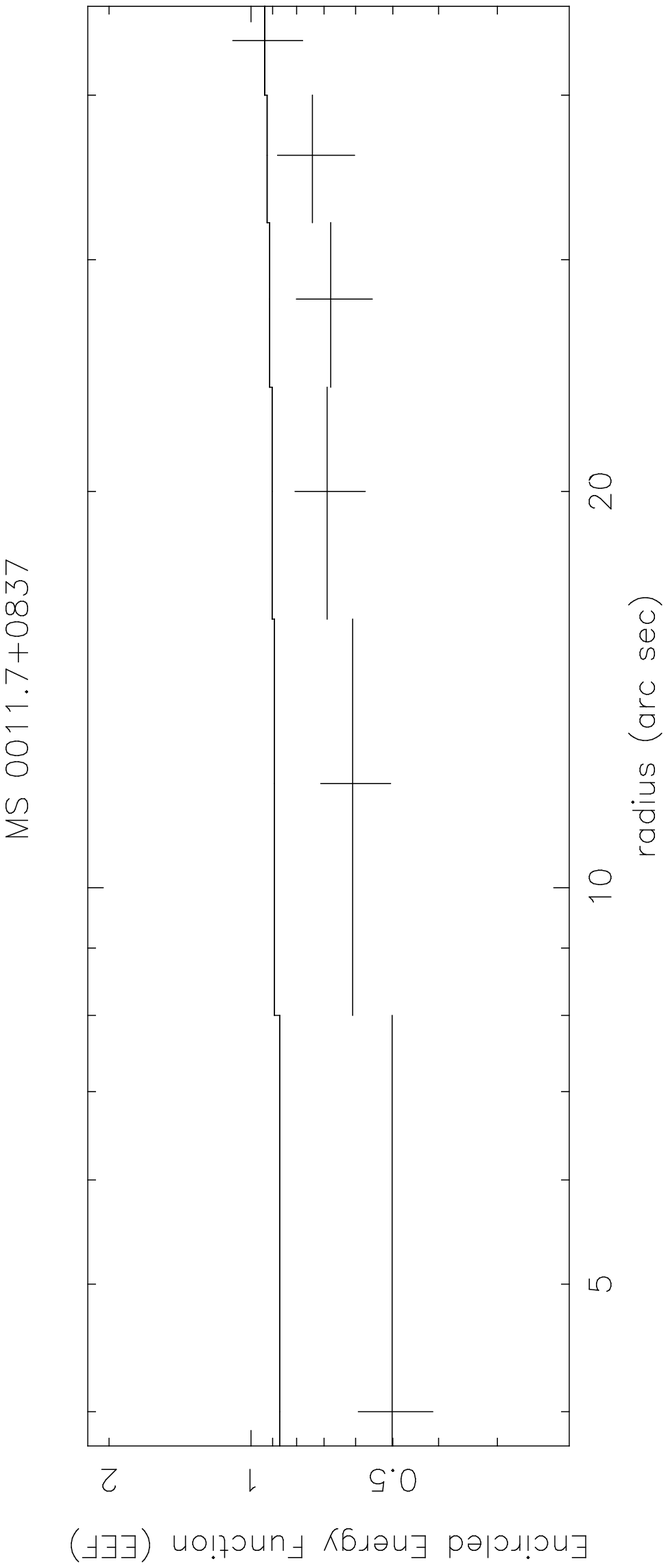}{0in}{-90.}{32.}{32.}{-125}{400}
\plotfiddle{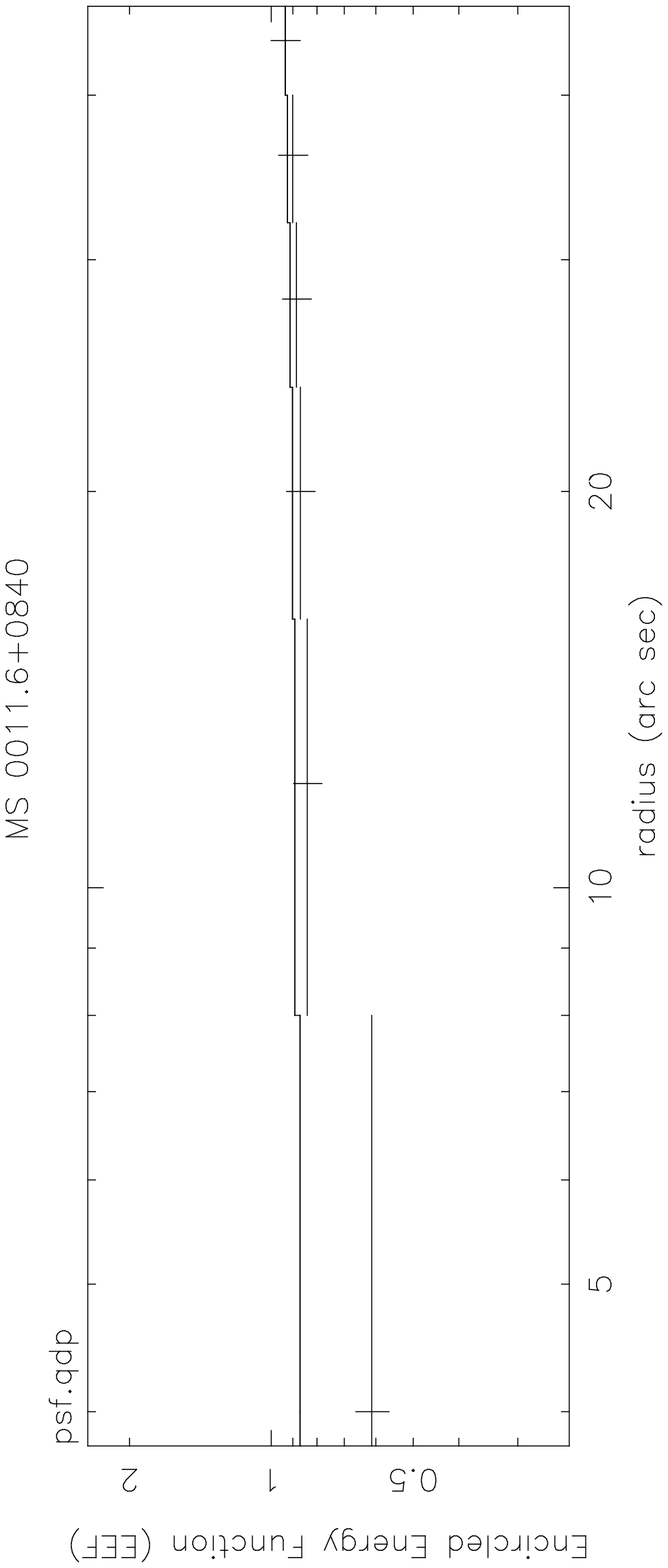}{0in}{-90.}{32.}{32.}{-125}{320}
\plotfiddle{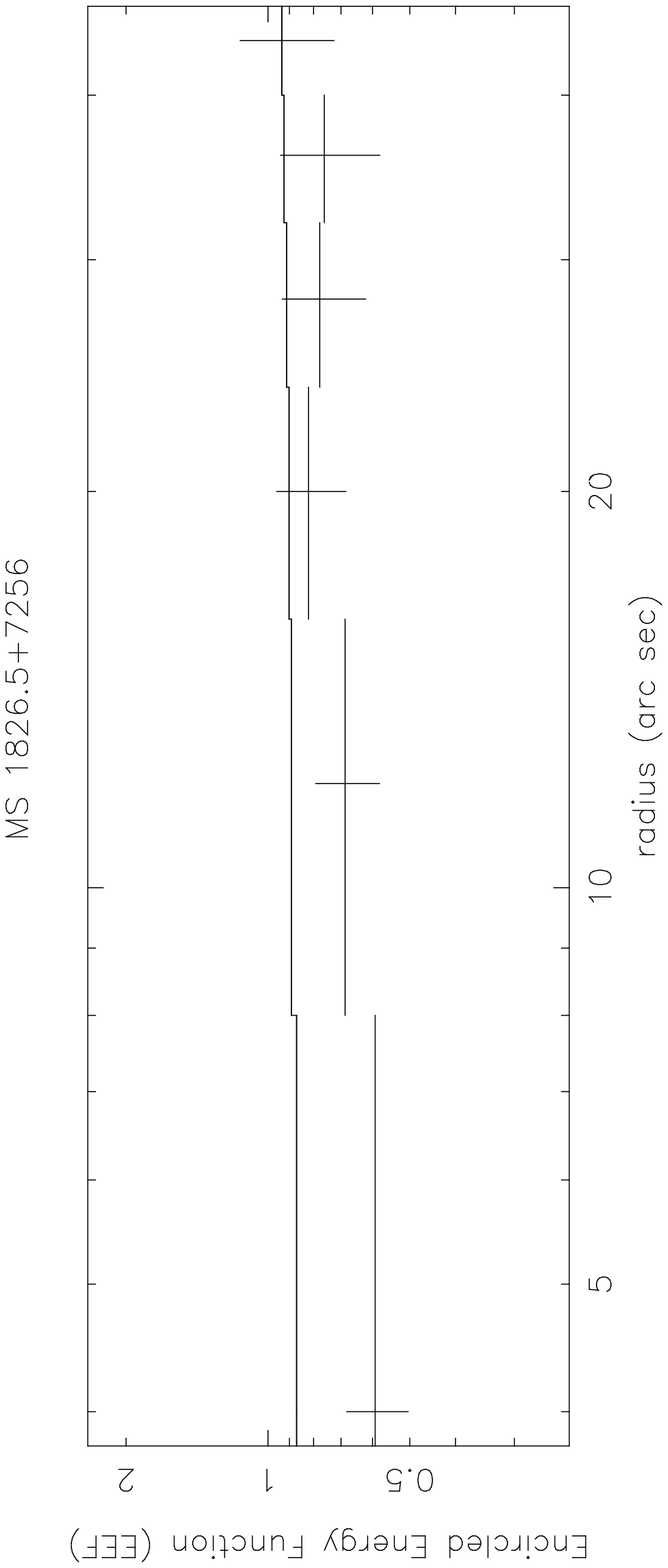}{0in}{-90.}{32.}{32.}{-125}{180}
\plotfiddle{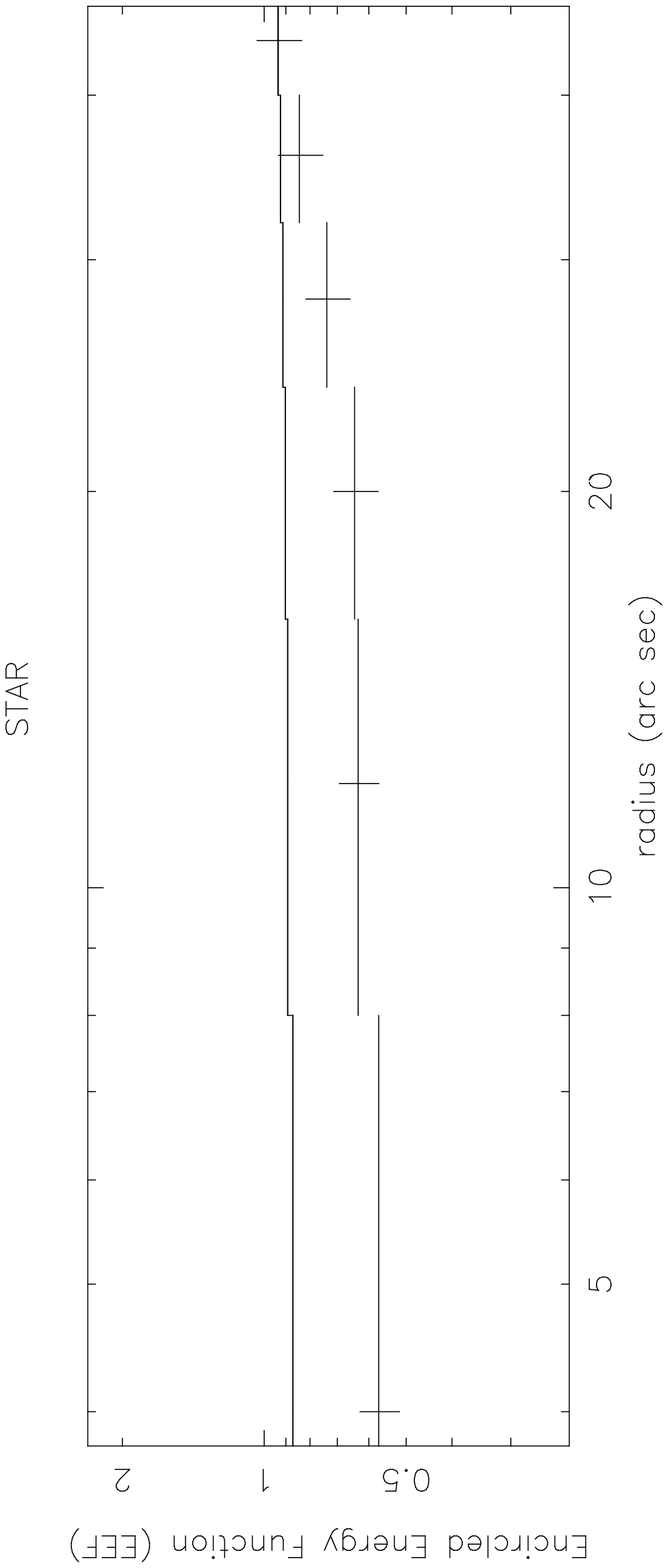}{0in}{-90.}{32.}{32.}{-125}{100}

\caption{The radial energy profiles for MS 0011.7+0837 and MS 1826.5+7256 as well as
for in-field stars plotted against the point-source EEF for the ROSAT HRI (see
Figure~\ref{fig1} for a full description of the EEF).  The radial energy profile
for MS 0011.7+0837 is not consistent with the nominal HRI EEF, whereas the
in-field star MS 0011.6+0840 does match the HRI EEF (excepting the ``excess
halo" described in \S 3).  The probability that the energy profile of MS
0011.7+0837 is the same as MS 0011.6+0840 is $\sim1$\% (the variance summed over all
the annuli is $\sigma = 3.77 \pm 0.27$).  In contrast, the radial energy profiles
for MS 1826.5+7256 and its in-field star are not consistent with the HRI EEF;
however, the radial energy profile for MS 1826.5+7256 is completely consistent with
that of the star; the probability is $> 95$\% that they are the same ($\sigma = 0.01
\pm 0.34$).    This suggests that MS 1826.5+7256 is unresolved, but that an as yet
unexplained problem with the observation exists.  Excessive jitter due to the
length of the exposure is a possible explanation.
\label{fig3}}
\end{figure}

\begin{figure}

\plottwo{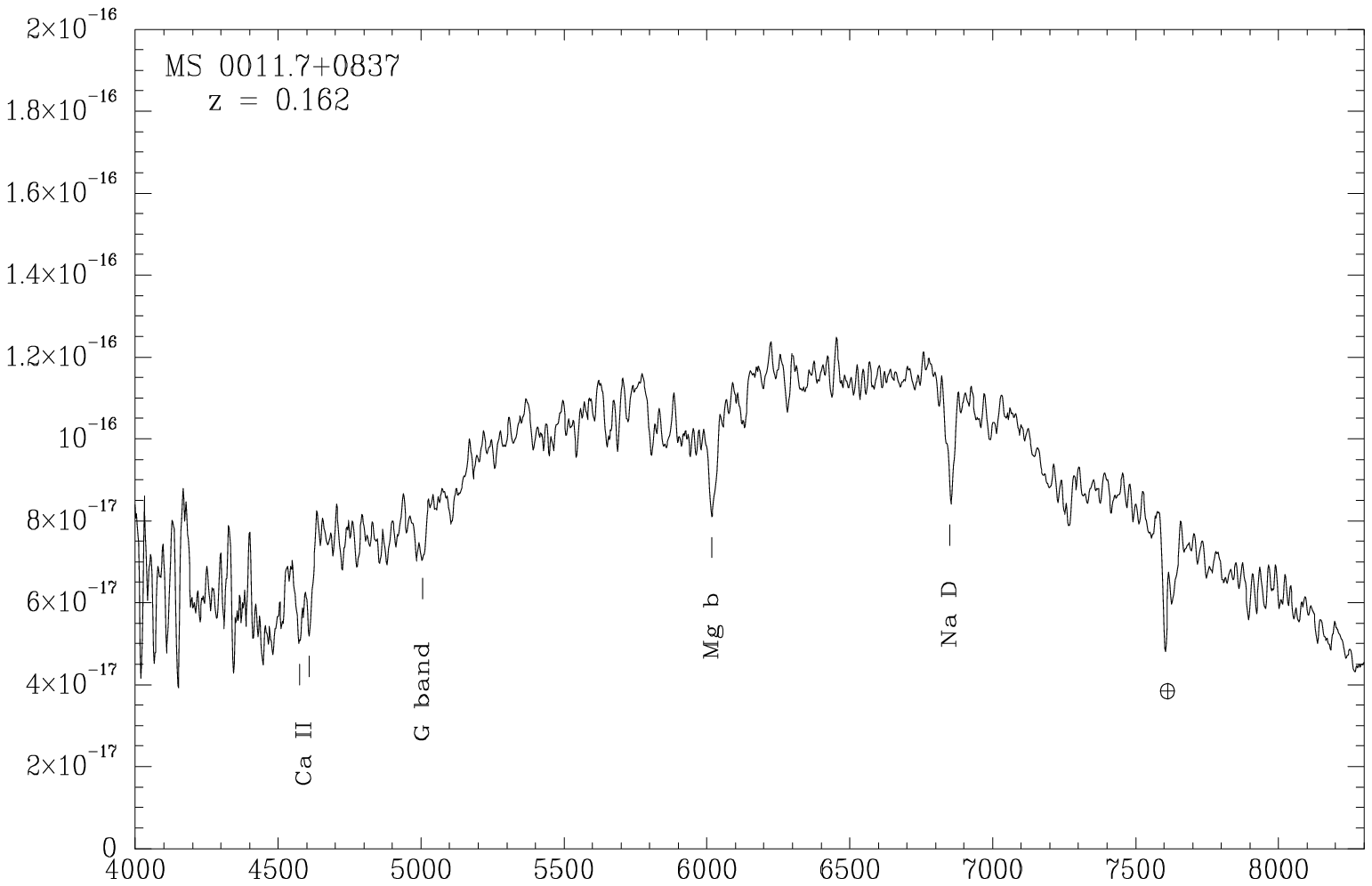}{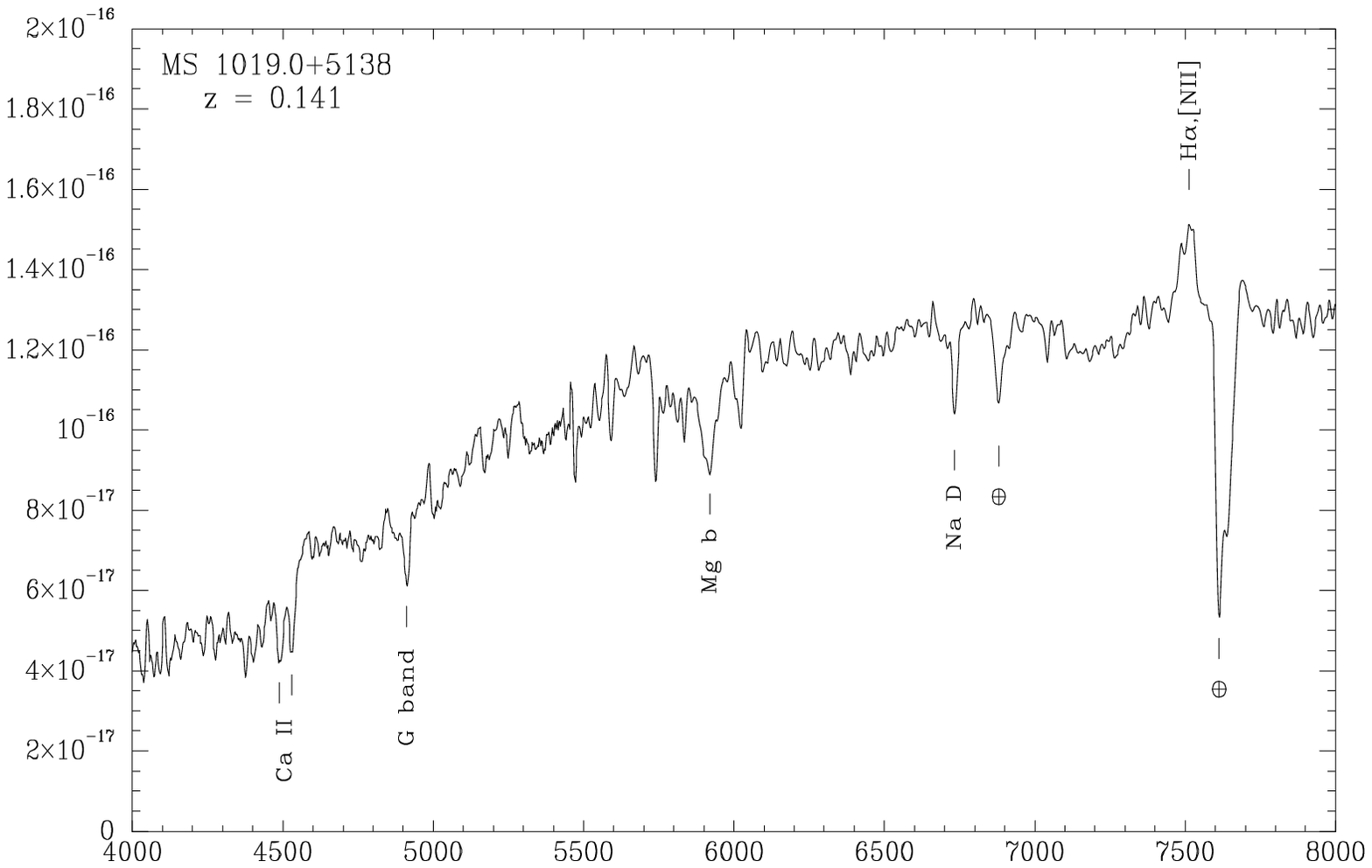}
\plottwo{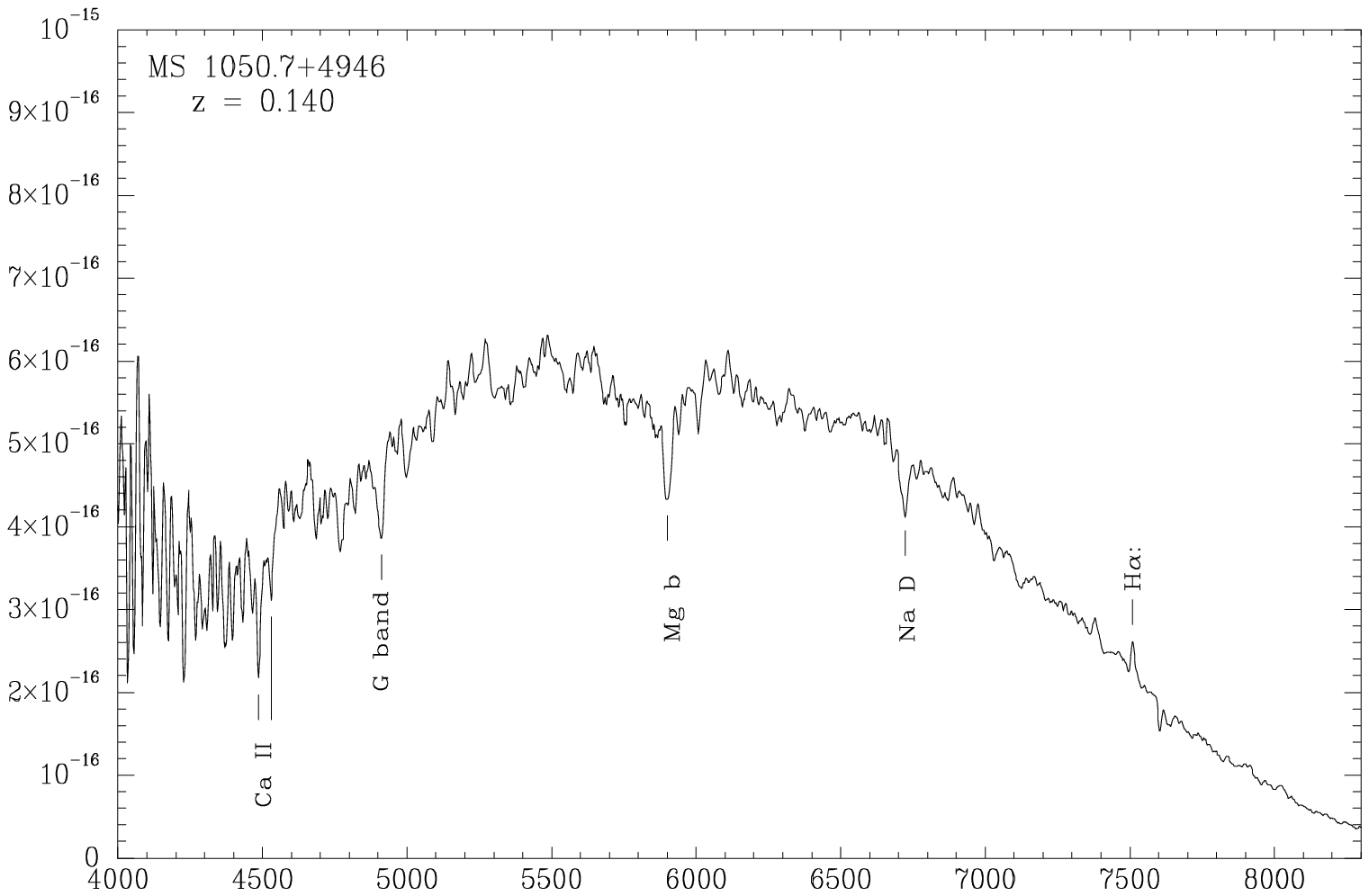}{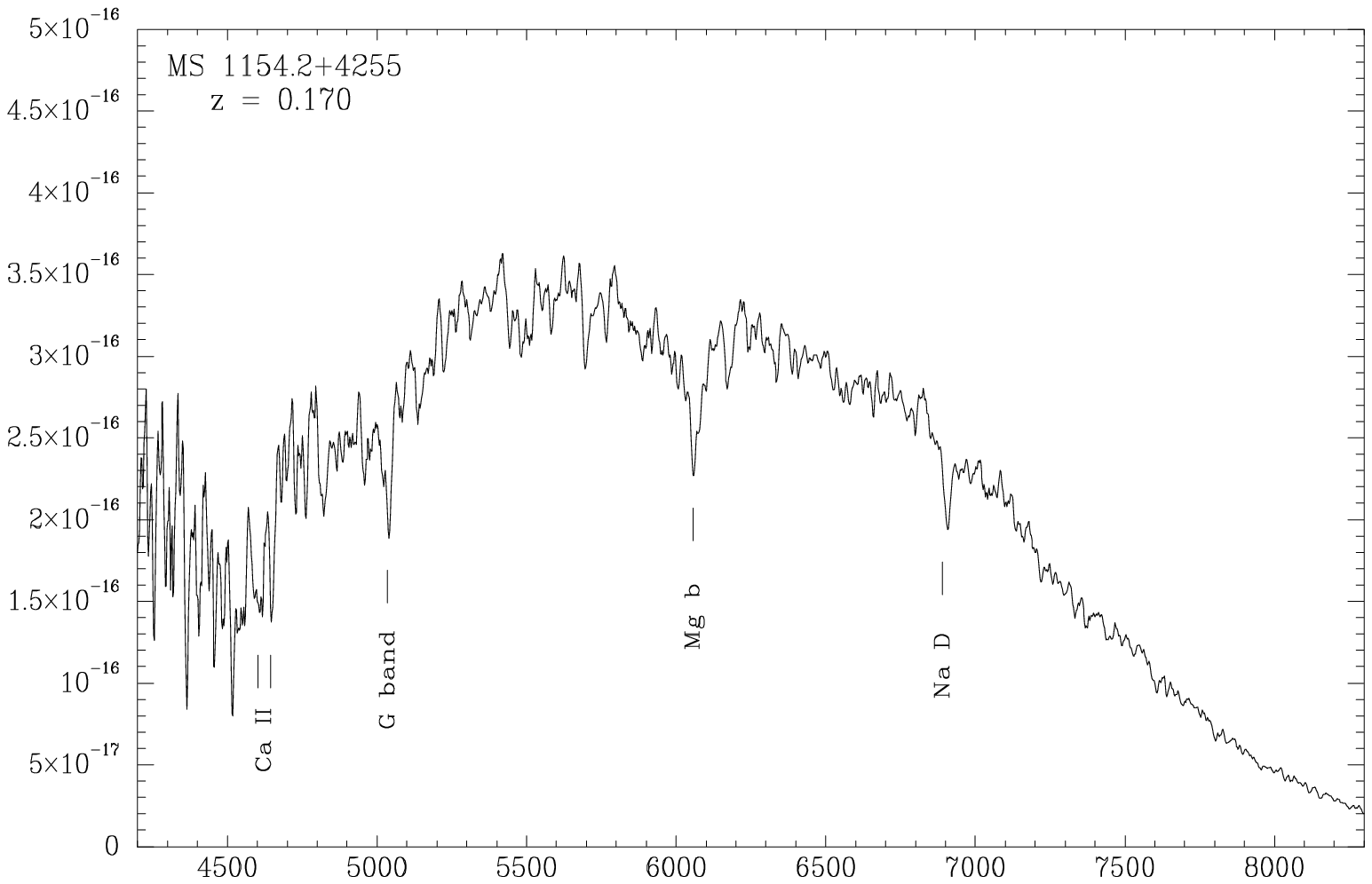}
\plottwo{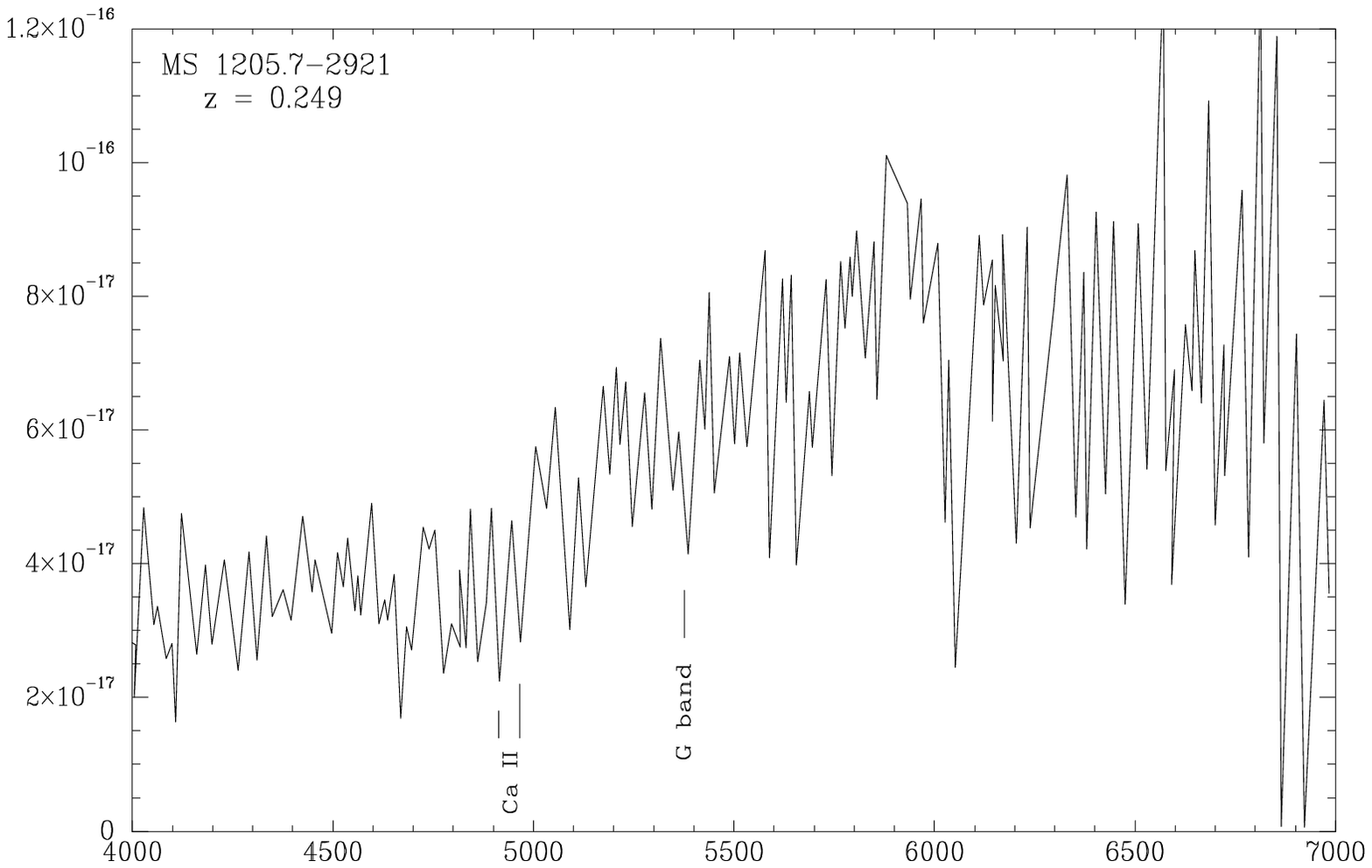}{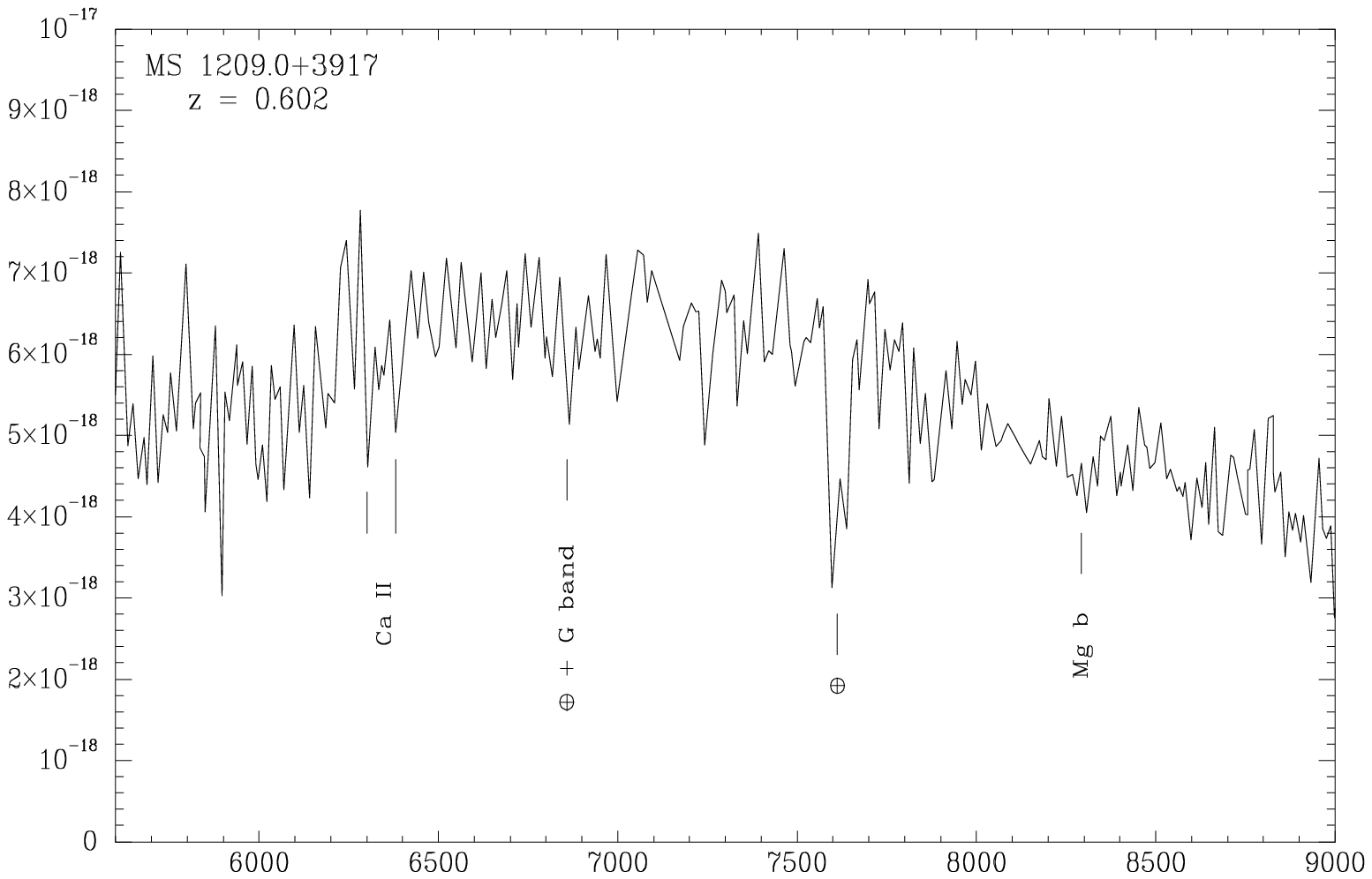}

\caption{Optical spectra of the BL Lac candidates and the radio galaxy in MS
0011.7+0837.  The flux scale is $F_{\lambda}$ in ergs s$^{-1}$ cm$^{-2}$; the
x-axis is wavelength in \AA.  The symbol $\earth$ identifies features due to the
Earth's atmosphere.  Due to the small slit width ($2\arcsec$) these fluxes are only
approximate, although the relative fluxes shortward of 6500\AA\ are reasonably
correct.  Longward of 6500\AA\ the flux calibration is incorrect due to
second-order overlap for all spectra except MS 1019.0+5138.}
\label{fig5}
\end{figure}

MS 0011.7+0837 is marginally resolved by the ROSAT HRI.  The peak of the X-ray
emission coincides with the core of a radio galaxy, but the X-ray emission is
elongated along the E-W axis (Figure~\ref{fig4}).  The maximum point source
contribution to the X-ray emission at the position of the radio galaxy is $\sim
50$\% of the observed ROSAT HRI flux (see Figure~\ref{fig3}); however this is $<
8$\% of the flux detected by the {\it Einstein} IPC (see discussion in \S 2).  Thus
this source was detected by the IPC because of the extended cluster emission, not
because of the radio galaxy.  The VLA A-array snapshot shows a classic FR-1
morphology (Figure~\ref{fig6}) with a very low core dominance value, typical of a
normal FR-1 radio galaxy.  The optical identification is a prominent cD radio
galaxy in a poor cluster at $\alpha = 00^h11^m45.5^s, \delta =
+08\arcdeg37\arcmin19.4\arcsec$ (B1950) in the image of Gioia \& Luppino (1994). 
The radio galaxy is the marked object in the finding chart for this source
(Maccacaro et al. 1994).  However, the Ca II break indicates the presence of a
non-thermal continuum very similar to the other low-luminosity BL Lacs identified
in this paper.  The above evidence suggests that MS 0011.7+0837 is a normal radio
galaxy or possibly a low-luminosity BL Lac within an X-ray cluster; however it is
the cluster that is responsible for the IPC detection (see \S 3).  The X-ray source
4.8\arcmin\ to the NW in Figure~\ref{fig4} is MS 0011.6+0840, the bright K0 V star
used in the point-source analysis of MS 0011.7+0837.  [ID = CLUSTER]

MS 0433.9+0957 is well resolved by the ROSAT HRI.  Although the X-ray emission
encompasses two radio galaxies, no distinct point sources are detected at their
locations; any contribution of X-ray flux from point-sources at their positions is
negligible. We identify this source as a cluster, in agreement with the optical
identification of a loose, poor cluster in Gioia \& Luppino (1994); one radio
galaxy is located at the center of their image, and the second is 30\arcsec\ south
and 20\arcsec\ east of center.  [ID = CLUSTER]

MS 1004.2+1238 is well resolved by the ROSAT HRI.  The optical field contains a
loose cluster dominated by a central radio-emitting galaxy (Stocke et. al. 1991). 
The radio and optical positions of the central galaxy do not match the HRI
position; the HRI position of the peak flux is $\sim 9\arcsec$ to the north of the
radio position (Figure~\ref{fig4}).  The maximum contribution of a point source at
the location of the radio galaxy is $\sim 20$\% the total X-ray flux.  No X-ray
flux is detected at the position of another radio galaxy $\sim 3$\arcmin\ north of
the cluster.  The ROSAT HRI and {\it Einstein} IPC fluxes match only to within
50\%, indicative of extended flux undetected by the HRI.  The radio galaxy is the
east of the two nuclei at the position marked by Maccacaro et al. (1994) in their
finding chart.  The optical spectrum of the radio galaxy has a strong CaII break 
($D(4000) = 1.9$) and strong [O II] emission ($W_{\lambda} = 40$\AA), further
suggesting it is not a BL Lac.  [ID = CLUSTER]

MS 1019.0+5139 is unresolved by the ROSAT HRI.  A deep B-array 20cm VLA
map does not detect any extended structure above the 0.02 mJy beam$^{-1}$
noise level.  Assuming a morphology and physical extent typical of XBLs
($\sim 100$ kpc; Perlman \& Stocke 1993) places a very low limit of log
$L_r < 22.67$ on its extended radio luminosity.  The absence of extended flux to
such a low-luminosity level is unusual for XBLs and is also rare among
RBLs.  Given its low redshift, a deep C-array 1.4 GHz mapping should be done to
look for very diffuse extended structure.  The radio core position matches the HRI
position on the object marked in the finder of Maccacaro et al (1994).  H$\alpha$
and [NII] 6549,6583\AA\ are detected in emission in this object at the same
redshift as the host galaxy; the
$W_{\lambda}$ for these lines 
are within the BL Lac classification limits.  Its CaII break strength $D(4000)
= 1.30$ (a ``contrast'' of 23\%) is also consistent with the BL Lac selection
criteria of Stocke et al. (1991). [ID = BL LAC]

MS 1050.7+4946 is unresolved by the ROSAT HRI.  The B-array 20cm VLA map shows a
classic FR-1 morphology but with a core to extended flux ratio quite large for a
typical FR-1 (see Figure~\ref{fig6} and Table~\ref{tbl-3}).  The optical field
shows a central dominant radio galaxy (marked in the finder chart of Maccacaro et
al. 1994) in a very sparse cluster typical for BL Lacs (Wurtz et al. 1997).  The
radio position matches the HRI position.  The optical spectrum shows possible weak
($W_{\lambda} = 2.5$\AA) H$\alpha$ in emission.  Its CaII break strength $D(4000) =
1.47$ (a ``contrast'' of 32\%) is somewhat strong for the selection criteria of
Stocke et al. (1991), but it is not inconsistent with the relaxed criteria proposed
by March\~a et al. (1996; see below); further, its X-ray and radio properties are
more consistent with a BL Lac than a normal FR-1 radio galaxy. [ID = BL LAC]

MS 1154.1+4255 is unresolved by the ROSAT HRI.  It also has the lowest X-ray
luminosity amongst the EMSS BL Lacs.  The radio position matches the HRI position. 
The optical field shows only a very sparse group of galaxies, with the BL
Lac marked in the finder of Maccacaro et al. (1994).  Like MS 1050.7+4946, its CaII
break strength $D(4000) = 1.5$ (a ``contrast'' of 33\%) is somewhat stronger than
allowed by the selection criteria; it appears transitional from BL Lacs to the
population of normal radio galaxies. The only available 20cm radio map for this
source is from the FIRST survey (White et al. 1997), where it is unresolved; it is
also unresolved in a VLA 6cm C-array snapshot (Gioia
\& Luppino 1994). [ID = BL LAC]

MS 1205.7-2921 is unresolved by the ROSAT HRI at the optical and radio position. 
This source was previously assigned a redshift of $z=0.171$ based upon a tentative
identification of Ca H\&K (Stocke et al. 1991).  The correct redshift is $z=0.249$
based upon the detection of a relatively strong CaII break ($D(4000) \approx 1.6$)
and G-band at this redshift (see Figure~\ref{fig5}).  The strength of the
CaII break is questioned due to the low quality of the available spectrum. 
Otherwise this object is mostly consistent with the BL Lac classification.  The
finding chart in Maccacaro et al (1994) identifies this BL Lac. [ID = BL LAC]

MS 1209.0+3917 is unresolved by the ROSAT HRI at the optical and radio position. 
This object was originally identified as a ``cooling flow galaxy''
(Stocke et al. 1991) based upon a poor optical spectrum, but it unambiguously meets
the BL Lac classification criteria based upon the new optical spectrum shown in
Figure~\ref{fig5}. 
The only 
available 20cm radio map for this
source is from the FIRST survey (White et al. 1997), where it is unresolved; it is
also unresolved in a VLA 6cm C-array snapshot (Gioia
\& Luppino 1994).
The radio position matches the HRI position on the object marked in the finder chart of Maccacaro et
al. (1994).  In the Gioia \& Luppino (1994) image the BL Lac is at an approximate
offset of 35\arcsec\ south and 17\arcsec\ east. [ID = BL LAC]

MS 1317.0-2111 is undetected by the ROSAT HRI.  No sources are detected {\it at all}
within the HRI field of view.  A 3$\sigma$ point-source detection limit at the
position of the optical identification or the position of the blank-field radio
source indicates its HRI flux is $f_x \leq$ 1.2 x 10$^{-13}$ erg cm$^{-2}$
s$^{-1}$.  Converted to the IPC band assuming $\alpha_x = -1$ the upper limit on the
flux is less than 30\% of that measured by the {\it Einstein} IPC. No extended
flux was detected within the IPC error circle; but we cannot exclude the possiblity
that the IPC source is very extended. The absence of an HRI detection is also
consistent with a highly variable point source such as a BL Lac.  It is also
possible that this source is spurious.  A deeper X-ray image is necessary to
determine the nature of this source. [ID = UNKNOWN]

MS 1333.3+1725 was  on our original observing list but  was instead targeted by M.
Hattori.  Hattori \& Morikawa (1998, private communication) report that this X-ray source is
point-like at the position of an elliptical galaxy with a depressed Ca II break
(I.M. Gioia 1998, private communication).  Therefore, this source is a BL Lac object, not a cluster
as originally reported by Stocke et al. (1991).  Because it is not a member of the
M91 complete sample and because we have not had the opportunity to scrutinize the
X-ray data in the manner herein, this source will be discussed at a later time
(Rector et al. 1999).

MS 1520.1+3002 is undetected by the ROSAT HRI at the position of the original
optical identification.  However, very extended X-ray emission is marginally
detected coincident with a bright elliptical galaxy $\sim 30$\arcsec\ south of the
IPC error circle in Maccacaro et al. (1994).  A weak ($f_x
\approx 5$ x 10$^{-14}$ erg cm$^{-2}$ s$^{-1}$), resolved X-ray source at $\alpha =
15^h20^m11.4^s, \delta = +30\arcdeg05\arcmin38.3\arcsec$ (B1950) is also marginally
detected 2\arcmin\ north of the original optical identification, which may have
caused the measured IPC position to be moved north.  The nature
of this source is unknown.  While no optical spectrum was taken of
the bright elliptical galaxy to the south, it is likely at the same redshift as
members of the loose cluster within the field at $z=0.117$.  [ID = CLUSTER]

MS 1826.5+7256 is unresolved by the ROSAT HRI. This object does not match the EEF
as well as the other unresolved sources but it does match the EEF extracted from an
in-field star (Figure~\ref{fig3}).  This circumstance is most likely due to jitter
during the long integration time.  The HRI source position coincides with the
optical position of an M5 V star.  The optical field shows only two galaxies in the
vicinity (Stocke et al. 1991), further supporting it is not a cluster. The marked
object in the finding chart of Maccacaro et al. (1994) is {\it not} the correct
identification; the M5 V star at $\alpha = 18^h26^m25.8^s, \delta =
+72\arcdeg56\arcmin06.5\arcsec$ (B1950) is the X-ray source, $\approx 20$\arcsec\
southwest of the marked object. [ID = STAR]

MS 2301.3+1506 is well resolved by the ROSAT HRI.  We identify it as a cluster, in
agreement with the optical identification of a rich cluster with a dominant cD
galaxy by Gioia \& Luppino (1994). The peak of the X-ray flux coincides with the
optical and radio position of a radio galaxy (Figure~\ref{fig4}); however the
maximum contribution of a point source at the location of the radio galaxy is
$\sim 20$\% of the total X-ray flux; thus the extended X-ray flux dominates over any
point source present.  The radio galaxy is centered in the image of Gioia \& Luppino
(1994) and is the marked object in the finding chart of Maccacaro et al. (1994).  A
possibly unresolved X-ray source is also detected 4.3\arcmin\ to the SW at
$\alpha = 23^h01^m03.1^s, \delta = +15\arcdeg04\arcmin04.8\arcsec$ (B1950), which is
well outside of the IPC error circle for MS 2301.3+1506 and does not contribute to
the X-ray emission detected by the IPC for this source.  This SW source was excluded
from the EMSS because it was only a 2.8$\sigma$ IPC detection; however, its HRI
flux is rather bright ($f_x = 6.6$ x $10^{-13}$ erg s$^{-1}$ cm$^{-2}$), suggesting
strong variability.  It is in the field of view of the $R$-band image of MS
2301.3+1506 taken by Gioia \& Luppino (1994); and its optical position matches a
potential cluster member.  Spectroscopy of this galaxy should be obtained to
determine the nature of this source. [ID = CLUSTER]

\section{Results and discussion}

Six of the twelve X-ray sources observed by the ROSAT HRI are unresolved at
$\sim 3$\arcsec\ resolution: five are identified as BL Lac objects and one is
identified as a M5 V star (MS 1826.5+7256).  Since only 5 of the best 12 cases for
identifications are actually BL Lacs, we have now identified the large
majority of the EMSS BL Lacs, particularly in the well-scrutinized M91 complete
sample.  Thus, while some BL Lacs were originally misidentified in the EMSS, there
are few such cases. And while there may still be a few BL Lacs embedded in ``cooling
flow'' clusters that have not been thoroughly scrutinized using high resolution X-ray imaging,
they are few in number and do not alter substantially the conclusions we draw in this Section.

\begin{figure}
\plotone{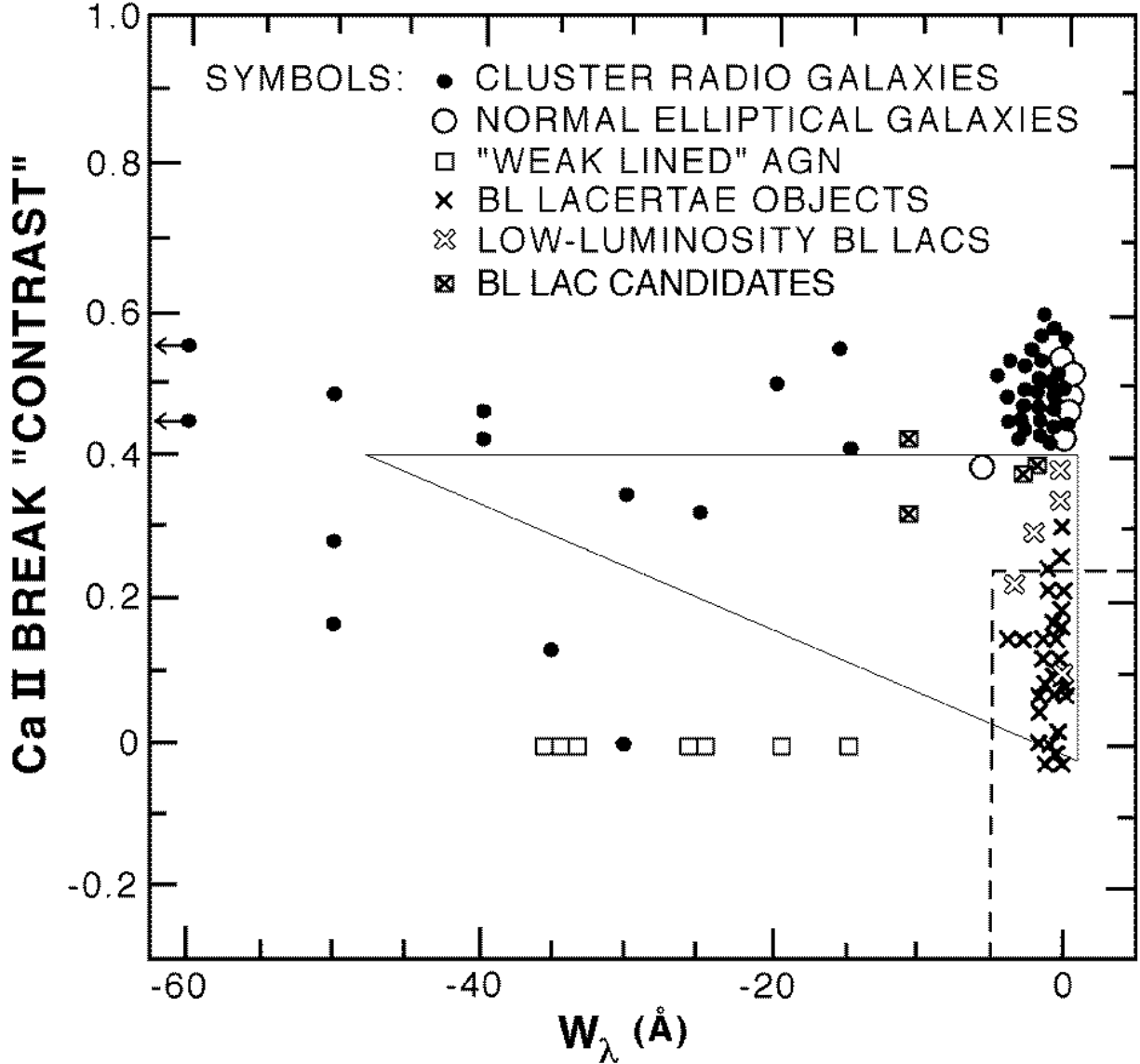}

\caption{CaII break ``contrast'' vs. equivalent width ($W_{\lambda}$) for a sampling
of extragalatic EMSS objects.  This plot (adapted from Stocke et al. 1991) uses the
largest $W_{\lambda}$ of any emission line present and the break ``contrast'' at
CaII H\&K (defined as the relative flux depression across the H\&K break 
($=[f(4000$\AA$^+)-f(4000$\AA$^-)/f(4000$\AA$^+)]$).  The subregion which meets the
BL Lac classification criteria suggested by Stocke et al. (1991) is shown in dashed
lines; i.e., $W_{\lambda} \leq 5$\AA\ and a break contrast $\leq 25$\% ($D(4000)
\leq 1.33$; Dressler \& Shectman 1987).  Note that three of the new
``low-luminosity" BL Lacs, along with the BL Lac MS 2306.1-2236, have stronger H\&K
contrasts than this criterion. These objects however do match the BL Lac criteria
proposed by March\~a et al. (1996; solid lines).  Three of the four BL Lac
candidates of Owen, Ledlow \& Keel (1996) also fall within this region; however we
note that very few BL Lac-like objects in the EMSS are detected within this
triangular region but outside the region considered herein for BL Lac objects.}
\label{fig7}
\end{figure}

\begin{figure}
\plottwo{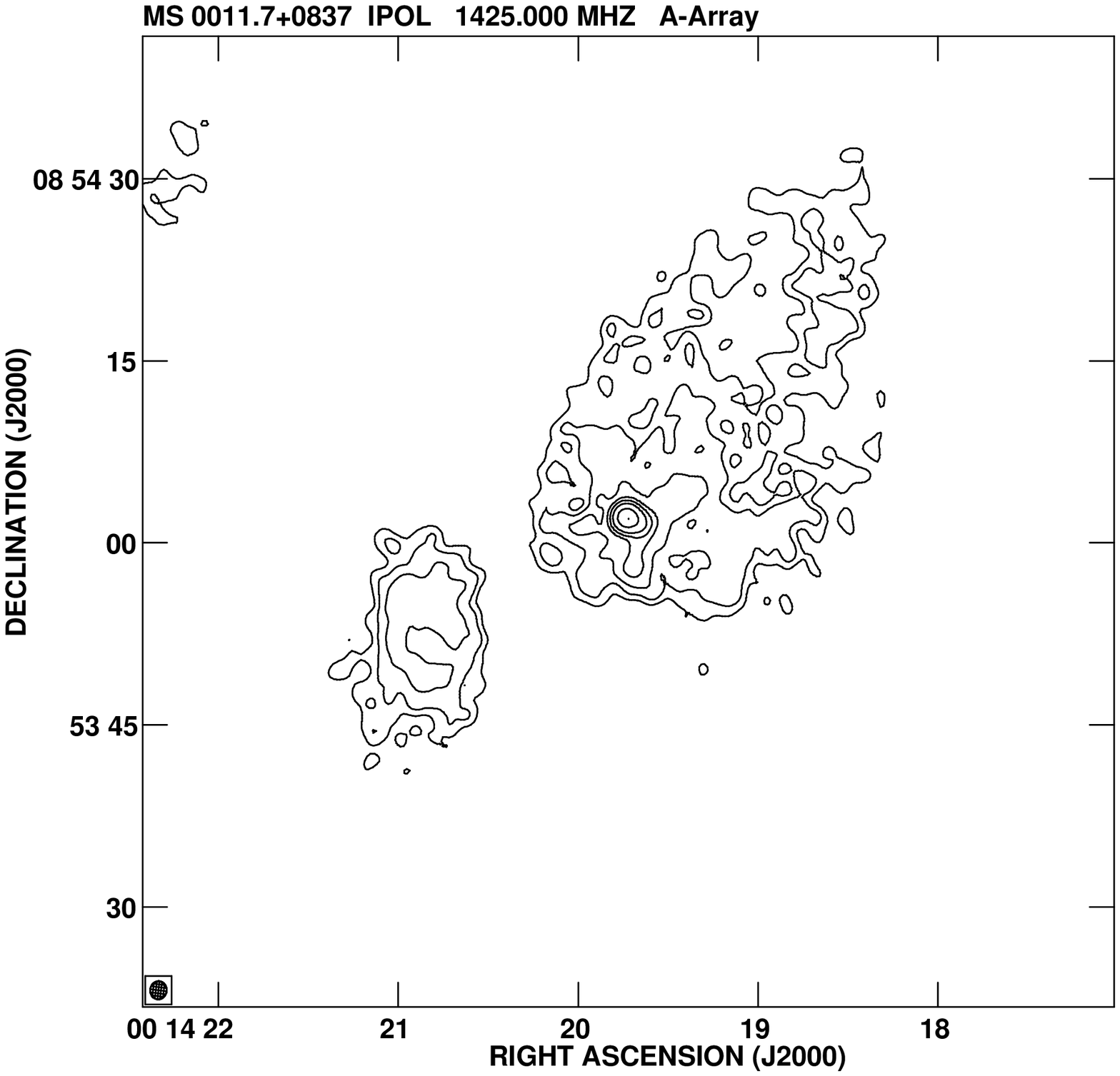}{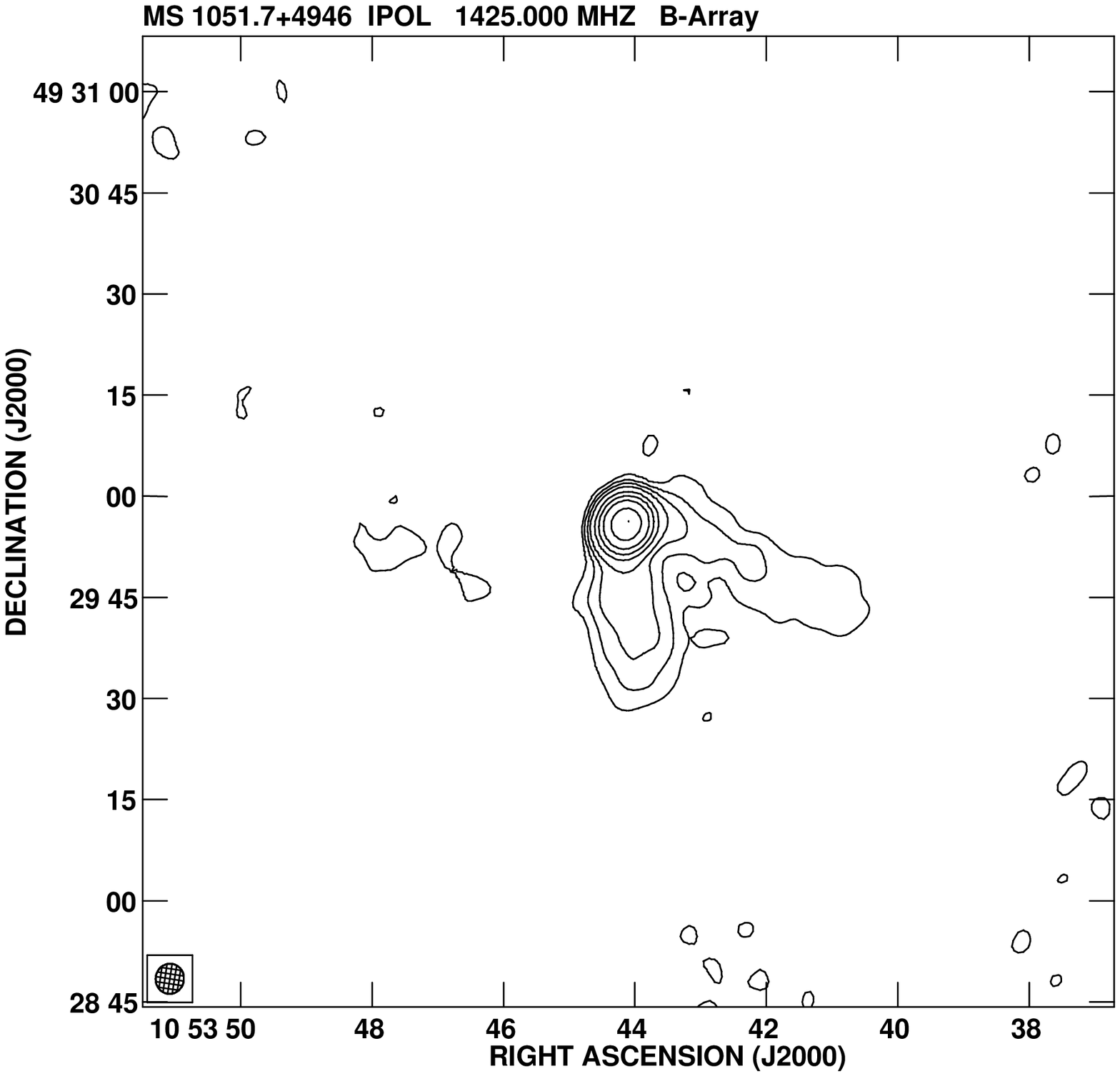}

\caption{VLA 20cm radio maps of MS 0011.7+0837 (A-array) and MS 1050.7+4946
(B-array).  The base level of each map is set roughly at the 2$\sigma$ RMS noise
level.  For MS 0011.7+0837 the peak flux is 47.8 mJy; and the contours are 0.5, 1,
2, 5, 10, 20, 50 and 100\% of the peak flux.  For MS 1050.7+4946 the peak flux is
49.4 mJy; and the contours are 0.2, 0.5, 1, 2, 5, 10, 20, 50 and 100\% of the peak
flux.  The beam is shown in the lower left.  A map of MS 1019.0+5139 is not shown as
extended flux was not detected for this source at 20cm in either A- or B-array.}
\label{fig6}
\end{figure}

\begin{figure}
\plotone{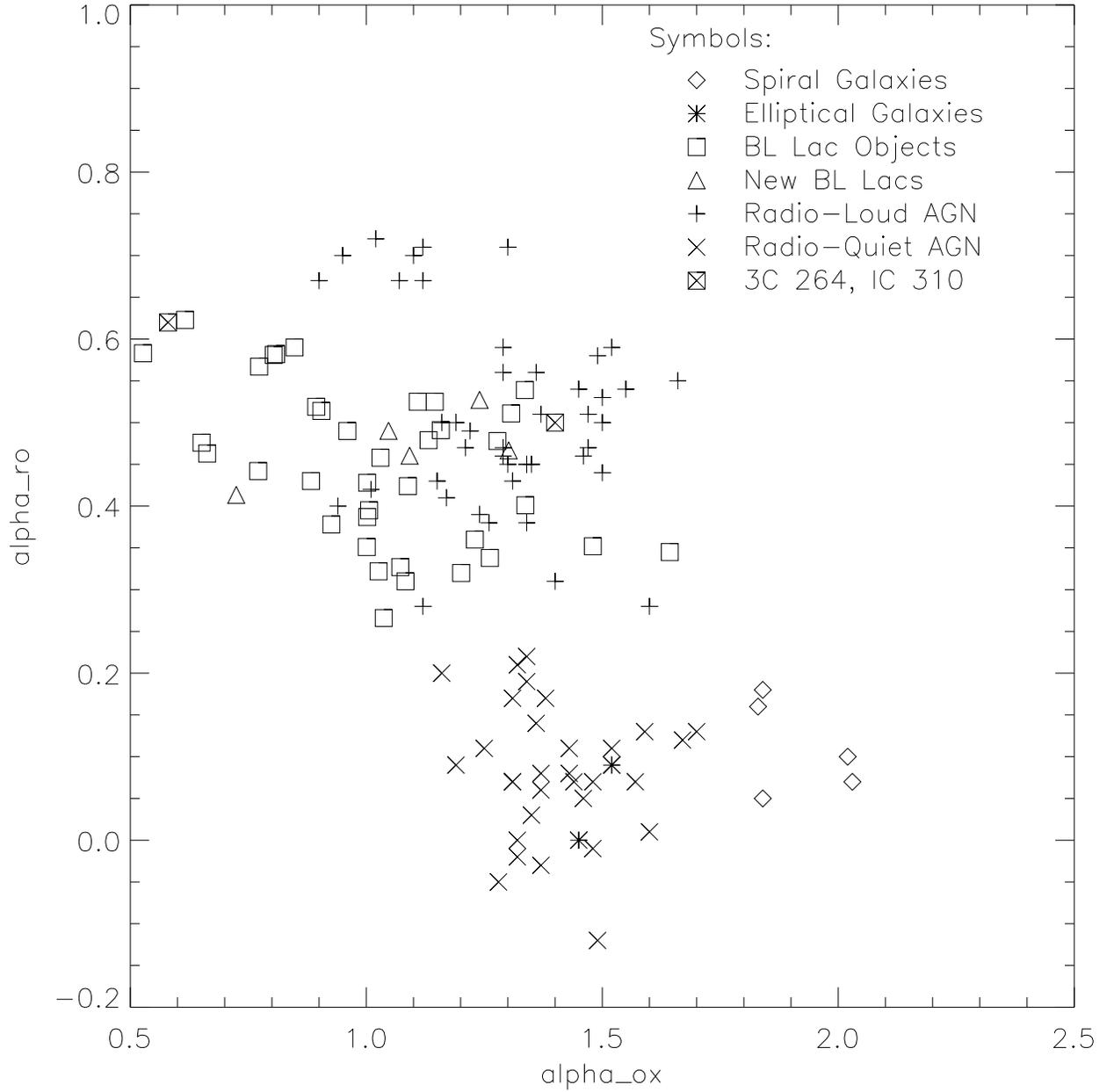}

\caption{Overall energy distributions for extragalactic sources in the EMSS which
were detected at 6cm (Stocke et al. 1991).  The axes are the radio-to-optical and
X-ray-to-optical spectral indices ($\alpha_{ro}$ and $\alpha_{ox}$ respectively) as
defined in Stocke et al. (1991).  Radio-loud objects are at the top of the plot and
X-ray-loud objects are to the left.   The newly discovered BL Lac objects (shown as
triangles) have similar ($\alpha_{ox},\alpha_{ro}$) values as compared to the other
EMSS BL Lacs.  The ($\alpha_{ox},\alpha_{ro}$) values for all of the EMSS BL Lacs as
well as the radio galaxies 3C 264 and IC 310 have been corrected to account for the
presence of the host elliptical galaxy (see \S 4 for a description).
Note that BL Lacs and radio-loud AGN lie in a region of this plot
relatively distinct from the radio-quiet AGN and normal elliptical galaxies.
}
\label{fig8}
\end{figure}

The optical spectra of these new BL Lacs are ``featureless,'' in the sense of
possessing no or only very weak emission lines.  However, the CaII break in some of
these sources is slightly stronger than in other XBLs (see Figure~\ref{fig7}), as
specifically predicted by BM93.  MS 1050.7+4946,  MS 1154.1+4255, MS 1205.7-2921 and
MS 2306.1-2236 are examples of low-$z$ objects which match the BL Lac classification
except for their slightly stronger CaII break strengths, pushing the limits of the
selection criteria of Stocke et al. (1991).  Their CaII break strengths are located
``in the gap'' between BL Lacs and FR-1 radio galaxies (see Figure~\ref{fig7}),
suggesting a smooth transition between the two populations.  The value of the CaII
break strength criterion of 25\% ($D(4000) \leq 1.33$) suggested by Stocke
et al. (1991) is somewhat arbitrary.  March\~a et al. (1996) note that nearly all
($> 95$\%) galaxies with a CaII break strength $\leq 40\%$ ($D(4000) \leq 1.67$)
must have an extra source of optical continuum.  These objects otherwise have BL
Lac-like characteristics including luminous point X-ray emission and low-power,
core-dominated radio sources.  Thus, we feel confident in classifying them as BL Lac objects.  

The broadband spectral energy distributions ($\alpha_{ox}, \alpha_{ro}$) for the new
BL Lac objects, once corrected for the presence of the luminous host elliptical
galaxy as described in \S 4, are similar to the other BL Lacs in the EMSS.  Thus,
as with the other EMSS BL Lac objects, the photometric classification scheme of
Stocke et al. (1991) confirms these objects to be BL Lacs.

Owen, Ledlow \& Keel (1996) discovered four FR-1 radio galaxies which they describe
as potential ``weak" BL Lac objects because they are BL Lac-like, with no strong
optical emission lines and a nonthermal origin of the depressed CaII break.  These
four objects do meet our relaxed CaII criteria but two of them, 3C 264 and IC 310,
show H$\alpha$+[N II] emission lines whose $W_{\lambda}$ slightly exceed the BL Lac
criterion.  However, we note that the $W_{\lambda}$ of these lines would be below
the criterion limit if the nonthermal optical continuum in these objects were
increased only slightly such that $D(4000) \leq 1.33$.  These objects do meet the
relaxed $W_{\lambda}$, $D(4000)$ criteria proposed by March\~a et al. (1996; see
Figure~\ref{fig7}).  Like MS 1019.0+5139, the strength of the H$\alpha$+[N II]
emission lines in 3C 264 and IC 310 are comparable to FR-1s (log
$L \approx 40$ erg s$^{-1}$).  

Owen, Ledlow \& Keel (1996) argue against a BL Lac identification for these four
objects because they are lobe-dominated radio sources ($f_{core} < f_{ext}$).  While
this is unusual for BL Lacs, it is consistent with being a source that is not highly
beamed.  The relationship of these objects to the low-luminosity BL Lacs discovered
here is not yet known, although they are similar in many ways.  ROSAT PSPC
observations are available for two of these four weak BL Lac candidates (3C 264 in
Abell 1367; Prieto 1996 and IC 310 in The Perseus Cluster; Rhee et al. 1994).  Both
of these AGN have apparent point-source detections which coincide with the radio
core, strongly suggestive that they are low-luminosity BL Lac objects.  3C 264 and
IC 310 have been placed in the ($\alpha_{ro},\alpha_{ox}$) plot in
Figure~\ref{fig8} by estimating the point-source X-ray fluxes and by correcting the
optical flux for the presence of the host elliptical.  Their location in the BL Lac
area confirms our inference that these objects are weak BL Lacs.  ROSAT HRI
point-source detections at the location of the other two radio galaxies (1005+006
and 1442+195) would confirm these to be BL Lac objects as well as provide
confirmation of an important new method for finding low-luminosity BL Lacs.  We
note that 3C 264 is a member of the 1Jy catalog (K\"uhr et al. 1981); thus, if it
is indeed a BL Lac the 1Jy sample of BL Lacs (Stickel et al. 1991) is incomplete
(MB95).  We investigate this issue further in another paper (Rector \& Stocke 1999).

Of the five newly discovered EMSS BL Lacs only two are in the complete M91 sample. 
The X-ray luminosities of these two new BL Lacs are the two lowest in the M91
sample, with an average luminosity (log $<$$L_x$$>$ = 44.10) somewhat lower than
the other BL Lacs in the M91 sample (log $<$$L_x$$>$ = 45.32).  A $t$-test shows the
difference in luminosity distributions to be significant (probability $>99$\%); so
a few low-luminosity BL Lacs were systematically misclassified as clusters of
galaxies in the original accounting of the EMSS.

Are these the ``low-luminosity" BL Lacs predicted by BM93?   BM93 estimate that the
probability of correctly identifying a BL Lac with a flux near the sample limit
drops dramatically with decreasing redshift because the weak nonthermal continuum
from the AGN will be dominated by the host galaxy starlight.  Optically this
object's spectrum could have too strong a CaII break to be classified as a BL Lac
using the criterion suggested by Stocke et al. (1991).  Based upon this optical
classification criterion BM93 estimate that a BL Lac in the EMSS sample at
$z < 0.2$ would have a probability of less than 30\% (and those at $z < 0.1$ have
less than 5\%) of being correctly identified as BL Lacs in the EMSS survey.  Thus,
it is not surprising that both of the newly identified BL Lacs in the M91 sample
are at a $z \sim 0.2$ and have CaII breaks which slightly exceed the BL Lac
criterion.  However, in their analysis BM93 assume an X-ray flux limit for the EMSS
sample of $S_x = 10^{-8}$ Jy ($f_x = 7.7 $ x $10^{-14}$ ergs s$^{-1}$ cm$^{-2}$ for
the IPC band), which is more than six times fainter than the cutoff for the the
complete M91 sample.  Due to its bright flux limit the M91 complete sample is
relatively immune to the problem of misidentification described in BM93.  MB95
estimate that 25\% of the full EMSS cluster sample (5 or 6 objects) were
misidentified; therefore it is not surprising that we have discovered only two
instances of misidentification in the M91 sample.  An analysis of a complete X-ray
sample with a much fainter flux limit ($S_x \leq 10^{-8}$ Jy) is necessary to
confirm the BM93 hypothesis.  Such an analysis is in progress using the fainter
sources in the EMSS (Rector et al. 1999).

Because of its moderate sensitivity and substantial sky coverage the EMSS is a
suitable survey for discovering low-luminosity BL Lacs.  These new objects allow the
XBL X-ray luminosity function (XLF) to be extended to slightly lower energies
($44.0 <$ log $L_x < 44.3$).  The number of objects discovered at these lower
energies (two: MS 1019.0+5139 and MS 1050.7+4946) is consistent with an
extrapolation of the previous XLF derived in M91; that is, we do not measure any
significant steepening or flattening of the XLF to lower luminosities (log $L_x
\leq 44.3$) than the previously published XLF in M91.  Including these two new
low-luminosity BL Lac objects slightly raises the \vvmax\ value for the M91 XBL
sample to $0.399 \pm 0.06$.  Since the latest value for the 1Jy radio-selected BL
Lac sample is \vvmax\ $= 0.614 \pm 0.05$ (Rector \& Stocke 1999), these newly
discovered BL Lacs do not ameliorate the discrepancy between XBLs and RBLs. 
Therefore, the possibility that these two selection techniques are sensitive to
finding different types of BL Lacs must still be considered.  

Because the EMSS is a ``serendipitous" survey, it is possible the M91 sample may
have been biased against previously known low-$z$, bright BL Lacs which were
targets of IPC observations.  However, the number of such objects in the 835
deg$^2$ sky coverage of the EMSS is likely quite small, $\sim 1$ object based upon
known space densities (M91, Perlman et al. 1996a).  A recent sample selected by the
RASS (Bade et al. 1998) confirms this effect as it finds three BL Lacs at $z < 0.1$
in a sky area rougly three times that surveyed by the EMSS; two of these BL Lacs
(Mkn 180 and I Zw 187) were previously known.  If either of these two objects were
included in the EMSS, the \vvmax\ would slightly decrease due to their high
flux.  Therefore, the XBL/RBL \vvmax\ discrepancy remains despite this selection
bias in the EMSS.  Further, the new RASS XBL samples of Bade et al. (1998) and
Giommi, Menna \& Padovani (1998) confirm the negative evolution of XBLs.  

A discussion of the impact of these observations on the cluster X-ray luminosity
function will be included in a future publication (Lewis, Stocke, Ellingson \&
Gioia 1999).

\section{Conclusions}

We have used the ROSAT HRI to search for BL Lac objects originally misidentified as
galaxies or clusters of galaxies within the EMSS sample.  Five new BL Lacs have
been identified by this process, two of which are new members of the M91 complete
subsample.  Therefore, such misidentifications do not significantly affect the
completeness of the BL Lac sample of M91.  Because only two of the
seven potential M91 subsample objects were found to be BL Lacs (and only 5 of 12
overall); and because these 12 were reobserved with the ROSAT HRI as the most likely
misdentifications we expect that there are very few remaining
misidentified sources in the EMSS.  It is also unlikely that many more
low-luminosity BL Lacs will be discovered within the EMSS as nearly all the
potential candidates have been investigated.  Thus we believe that contamination of
the EMSS sample from misidentification is not significant.   

While it was the intention of this observing program to obtain ROSAT HRI images of all
suspect X-ray sources in the M91 subsample, we had not considered the possibility (raised by
the referee) that ``cooling flow'' clusters might hide point X-ray sources in their
central surface brightness ``cusps.''  We had not made this consideration heretofore because
this is not the usual interpretation of these cusps (Fabian 1994) and because the BCGs/radio galaxies
in these clusters contain luminous emission lines in their optical spectra which easily exceed
the equivalent width limits proposed for BL Lacs in Stocke et al. (1991). However, these
luminous, low ionization emission lines are also very spatially extended and so not obviously
related to an AGN broad- or narrow-line region. Once we reconsidered these cooling flow
clusters, most were found to be too extended to harbor point sources luminous enough to be
members of the EMSS sample on their own; but four clusters could not be so eliminated and so
remain as potential (but not proven to be) BL Lac objects. These are MS 0102.3+3255, MS
1244.2+7114, MS 1455.0+2232 \& MS 2348.0+2913. So, while it was the intention of this study to
complete the M91 BL Lac sample, these four clusters were not observed with the HRI and so
the M91 BL Lac sample may, indeed, contain a few more low-luminosity BL Lacs. 

Nevertheless, our results generally agree with the predictions of BM93.  The properties of the
low-luminosity BL Lacs we have discovered are consistent with the low-luminosity BL
Lacs predicted.  And at the relatively bright flux limit of the M91 sample,
MB95 predict very few low-luminosity BL Lacs would be misidentified, as we have
found.  Thus the B-M effect does not explain the \vvmax\ discrepancy between the
M91 EMSS sample and 1Jy BL Lac sample, even if all four of the cooling flow clusters are
included in the M91 BL Lac sample. 

The properties of the newly discovered BL Lacs are suggestive of a transition
population between BL Lacs and FR-1 radio galaxies, however their properties are too
similar to that of the other BL Lacs in the EMSS sample to completely ``bridge the
gap" between XBLs and standard FR-1s in all their properties.  But based upon this
small study, surveys with significantly larger sky area and similar sensitivity
(e.g., the RASS) will likely find numerous examples of low-luminosity BL Lacs whose
properties are continuous between luminous X-ray selected BL Lacs, as found in the
EMSS and Slew surveys, and normal FR-1 radio galaxies like M87.  Indeed, the
RGB (Laurent-Muelheisen et al. 1999) and Deep X-Ray Radio Blazar Survey (DXRBS; Perlman et al. 1998) surveys
have found significant numbers of BL Lacs within this gap.  These fainter
surveys are also more susceptible to the problem of recognition of low-luminosity BL
Lacs at low redshift, thus allowing for a more definitive test of the B-M effect.

Finally, the confirmation of IC 310 and 3C 264 as low-luminosity BL Lac objects
suggests that other, nearby X-ray luminous clusters could harbor low-luminosity BL
Lacs (similar to the case presented here for MS 0011.7+0837).  Since many X-ray
luminous clusters at $z < 0.1$ were targeted by the {\it Einstein} IPC, the EMSS
would not have included these sources.  However, the spectral results of Owen,
Ledlow \& Keel (1996), which included enormous numbers of nearby Abell cluster radio
galaxies, suggests that the numbers of low-luminosity BL Lacs in rich clusters is
quite small.  Still, ROSAT HRI images of suspect radio galaxies in nearby rich
clusters, particularly the ``cooling flow'' clusters,  should be targeted as potential 
BL Lac objects (e.g., 1005+006 \&
1442+195 from the Owen et al. list).  Similarly, HRI (or AXAF HRC) observations should be 
obtained  for the five $z < 0.2$
BL Lac candidates discovered in March\~a et al. (1996) which exceed the
$W_{\lambda}$,
$D(4000)$ criteria of Stocke et al. (1991).  Their ($\alpha_{ox}$ and
$\alpha_{ro}$) values, once corrected for the presence of the host galaxy, could
confirm their classification as BL Lac objects, thus accounting for the marked
absence of BL Lacs at $z < 0.2$.

\acknowledgments

This work was supported by NASA ROSAT grant NAG5-4518 and NASA LTSA grant NAGW-2645.
The authors thank Simon Morris for the use of the optical spectrum of MS 1205.7-2921
and Isabella Gioia for the use of the optical spectrum of MS 1209.0+3917.  We
also thank Kohji Morikawa and Makoto Hattori for communicating their results on
MS 1333.3+1725 prior to publication. We thank the referee for his or her intriguing
idea that ``cooling flow'' clusters could harbor BL Lac objects.
This research has made use of data obtained
through the High Energy Astrophysics Science Archive Research Center Online Service,
provided by the NASA/Goddard Space Flight Center as well as NASA's Astrophysics
Data System Abstract Service.

\clearpage
\begin{deluxetable}{llrrrrr}
\tablecaption{Log of ROSAT HRI Observations \label{tbl-1}}
\tablewidth{0pt}
\tablehead{
\colhead{Object} & \colhead{ROR\#} & \colhead{Date} &
\colhead{Exposure} & \colhead{M91?} & \colhead{RA (B1950)\tablenotemark{a}} &
\colhead{Dec}}
\startdata
MS 0011.7+0837 & RH701839 & 4  Jan 1995 & 6848  & Y & 00:11:45.3 & +08:37:24.3 \nl
MS 0433.9+0957 & RH701840 & 3  Mar 1995 & 7530  & Y & 04:33:57.2 & +09:56:48.3 \nl
MS 1004.2+1238 & RH702864 & 16 May 1997 & 7374  & Y & 10:04:12.5 & +12:38:22.1 \nl
MS 1019.0+5139 & RH600125 & 26 Oct 1991 & 7190  & Y & 10:19:03.5 & +51:39:09.1 \nl
MS 1050.7+4946 & RH701841 & 2  Dec 1994 & 2895  & Y & 10:50:47.6 & +49:45:51.7 \nl
MS 1154.1+4255 & RH701843 & 21 May 1995 & 6243  & N & 11:54:11.6 & +42:54:52.1 \nl
MS 1205.7-2921 & RH702867 & 3  Jan 1997 & 10651 & N & 12:05:43.2 & -29:21:18.3 \nl
MS 1209.0+3917 & RH701844 & 18 May 1995 & 11355 & N & 12:09:02.9 & +39:17:35.9 \nl
MS 1317.0-2111 & RH702865 & 1  Aug 1997 & 8491  & N & \nodata  &
\nodata\tablenotemark{b} \nl     MS 1520.1+3002 & RH702866 & 15 Jul 1997 & 10549 &
N & 15:20:11.4 & +30:05:38.3\tablenotemark{c}\nl MS 1826.5+7256 & RH600124 & 12 Sep
1991 & 36803 & Y & 18:26:25.1 & +72:56:10.2 \nl MS 2301.3+1506 & RH701842 & 20 Dec
1994 & 7287  & Y & 23:01:16.7 & +15:06:53.2 \nl
\tablenotetext{a}{For extended sources, the position marks the peak flux.}
\tablenotetext{b}{No sources are detected within the HRI field.}
\tablenotetext{c}{Position of northern source.  Does not coincide with
radio source (see discussion of this source in \S 5).}
\enddata
\end{deluxetable}

\clearpage
\begin{deluxetable}{rrrrrrrrr}
\tablecaption{X-ray Properties of BL Lac Candidates \label{tbl-2}}
\tablewidth{0pt}
\tablehead{
\colhead{Object} & \colhead{M91?} & \colhead{log $nH$} & \colhead{$\alpha_x$} &
\colhead{$f_{HRI}$} & \colhead{Log $L_{HRI}$} & \colhead{$f_{IPC}$} &
\colhead{$f_{IPC}$} &
\colhead{\%\tablenotemark{a}}  \\
\colhead{} & \colhead{} & \colhead{(cm$^{-2}$)} & \colhead{} &
\colhead{} & \colhead{(erg s$^{-1}$)} &
\colhead{($\alpha_x = 1$)} & \colhead{($\alpha_x$ = Col. 4)} & \colhead{} 
}
\startdata
0011.7+0837 & \nodata\tablenotemark{b} & 20.79 & 1.73 & 6.71 & 43.94 & 13.11 &
16.46 & 17\%  \nl  1019.0+5139 & Y & 19.98 & 0.56 & 47.8 & 44.67 & 13.28 & 17.06 &
298\% \nl  1050.7+4946 & Y & 20.11 & 1.62 & 26.3 & 44.40 & 14.09 &  8.86 & 136\%
\nl  1154.2+4255 & N & 20.15 & 1.68 & 5.35 & 43.89 &  4.04 &  2.72 & 86\%  \nl 
1205.7-2921 & N & 20.76 & 1.48 & 21.4 & 44.86 &  3.23 &  3.78 & 296\% \nl 
1209.0+3917 & N & 20.32 & 0.74 & 5.47 & 45.15 &  2.83 &  2.39 & 146\% \nl 
\tablenotetext{a}{Ratio of ROSAT HRI to {\it Einstein} IPC flux.  See text for
a full description.}
\tablenotetext{b}{This object is not identified as a BL Lac (see text).}
\enddata


 
\end{deluxetable}

\clearpage
\begin{deluxetable}{lcrrcrrrrrr}
\tablecaption{Optical and Radio Properties of BL Lac Candidates \label{tbl-3}}
\tablewidth{0pt}
\tablehead{
\colhead{Object} & \colhead{$z$} & \colhead{$V$} & \colhead{$W_{\lambda}$} &
\colhead{$D(4000)$} &
\colhead{$S_{core}$} & \colhead{$S_{ext}$} & \colhead{$S_{tot}$} & \colhead{$f$} & \colhead{log
$L_{core}$} & \colhead{log $L_{ext}$} 
}
\startdata
0011.7+0837\tablenotemark{a} & 0.162 & 17.50 & $\leq2.1$ & $1.43\pm0.17$ &    47.1 &        219 &  266 &       0.21 &   24.75 &        25.35 \nl
1019.0+5139\tablenotemark{b} & 0.141 & 18.09 &       3.3 & $1.30\pm0.15$ &     3.1 & $\leq 0.6$ &  3.1 & $\geq 5.3$ &   23.44 & $\leq 22.67$ \nl  
1050.7+4946\tablenotemark{b} & 0.140 & 16.86 &       2.5 & $1.47\pm0.09$ &    52.8 &       13.2 & 66.0 &        4.0 &   24.67 &        24.01 \nl 
1154.2+4255\tablenotemark{c} & 0.170 & 17.67 & $\leq3.0$ & $1.50\pm0.23$ &    12.6 &        4.8 & 17.4 &        2.6 &   24.22 &        23.73 \nl 
1205.7-2921\tablenotemark{d} & 0.249 & 17.50 & $\leq4.0$ & $1.64\pm0.45$ & \nodata &    \nodata &  4.4 &    \nodata & \nodata &      \nodata \nl 
1209.0+3917\tablenotemark{c} & 0.616 & 20.00 & $\leq4.5$ & $1.17\pm0.17$ &    12.1 & $\leq 2.0$ & 12.1 &  $\geq 30$ &   25.39 &        24.38 \nl
\tablenotetext{a}{VLA 20cm A-array (mJy)}
\tablenotetext{b}{VLA 20cm B-array (mJy)}
\tablenotetext{c}{VLA 20cm B-array (mJy; FIRST survey; White et al. 1997)}
\tablenotetext{d}{VLA  6cm C-array (mJy; Stocke et al. 1991)}
\enddata
\end{deluxetable}

\clearpage

%

\clearpage

\clearpage

\clearpage

\clearpage

\clearpage

\clearpage

\clearpage

\end{document}